\documentclass[10pt]{article}

\usepackage[authoryear,round]{natbib}
\usepackage[margin=1in]{geometry}
\usepackage[T1]{fontenc}
\usepackage[utf8]{inputenc}
\usepackage{amsfonts}
\usepackage{amsmath}
\usepackage{booktabs}
\usepackage{changepage}
\usepackage{csvsimple}
\usepackage{enumitem}
\usepackage{float}
\usepackage{hyperref}
\usepackage{lmodern}
\usepackage{multicol}
\usepackage{needspace}
\usepackage{pgfplots}
\usepackage{pgfplotstable}
\usepackage{soul}
\usepackage{tabularx}
\usepackage{tikz}
\usepackage{url}
\usepackage{xstring}

\usetikzlibrary{arrows.meta, positioning}
\usetikzlibrary{shapes, arrows.meta, positioning, decorations.pathreplacing, calc, fit}

\pgfplotsset{compat=1.18}

\sethlcolor{yellow!60}

\title{Sift or Get Off the PoC: Applying Information Retrieval to Vulnerability Research with SiftRank}

\author{Caleb Gross \\ \texttt{c@leb.email}}

\date{\today}

\begin{document}
\maketitle

\begin{abstract}
	Security research is fundamentally a problem of resource constraint and
	consequent prioritization.  There is simply too much attack surface and
	too little time and energy
  to spend analyzing it all.
	The most effective security researchers are often those who are most
  skilled at intuitively deciding \textit{which} part of an expansive attack
  surface to investigate.
	We demonstrate that this problem---more generally, the problem of
  selecting the most promising option from among many possibilities---can be
  reframed as an information retrieval
	problem, and solved using document ranking techniques with large
  language models performing the heavy lifting as general-purpose rankers.

	We present SiftRank, a ranking algorithm achieving $O(n)$ complexity
	through three key mechanisms: listwise ranking using an LLM to order
	documents in small batches of approximately 10 items at a time;
	inflection-based convergence detection that adaptively terminates ranking
	when score distributions have stabilized; and
	iterative refinement that progressively focuses ranking effort on the most
	relevant documents. Unlike existing reranking approaches that require a
	separate first-stage retrieval step to narrow datasets to approximately 100
	candidates, SiftRank operates directly on thousands of items, with each
	document evaluated across multiple randomized batches to mitigate
	inconsistent judgments by an LLM.

	We demonstrate practical effectiveness on N-day vulnerability analysis,
	successfully identifying a vulnerability-fixing function among 2,197
	changed functions in a stripped binary firmware patch within
	99 seconds at an inference cost of \$0.82.
	Our approach enables scalable security prioritization for problems that 
  are generally constrained by manual analysis,
	requiring only standard LLM API access (including small models) without 
  specialized
	infrastructure, embedding, or domain-specific fine-tuning. An open-source 
implementation of SiftRank may be found at 
\url{https://github.com/noperator/siftrank}.
\end{abstract}

\section{Introduction}

The primary bottleneck in security research is not simply limited bug-detecting
capability---it is deciding \textit{which} attack surface to examine while
under severe resource constraints.
For a security practitioner, this manifests as questions like: Which code
scanner finding to triage? Which SOC alert to investigate? Which credentials to
use in a brute-force attack? Which data structure to fuzz? Which web app
injection point to test?
These challenges share a common structure: an overwhelming number of options to
choose from, subjective decision criteria that are difficult to quantify, and
finite resources to spend on evaluation.

We experienced this problem acutely while performing N-day vulnerability
analysis. When dissecting a firmware
patch\footnote{\url{https://cve-north-stars.github.io/docs/Patch-Diffing}} to
identify and reverse engineer a fixed authentication bypass vulnerability, we
faced the challenge of analyzing over 2,000 changed functions in a stripped
binary.
We came to recognize this problem as fundamentally one of \textit{ranking}.
That is, in this large but finite list of changed functions, which function
most likely fixes the vulnerability vaguely described in the security advisory?

Working from first principles to solve this practical problem, we initially
attempted using a large language model (LLM) to examine each changed function in the patch diff and
assign an advisory-relevance score that could be easily sorted. However, we
found two issues with this approach. LLMs struggle to produce a consistent
numerical score (even when provided with a clear scoring rubric), and important
detail is lost when compressing many dimensions of nuanced function data into a
single calibrated number. Since LLMs are capable of dealing with abstract
concepts and analyzing unstructured data, we considered instead: Why not simply
compare each function definition \textit{directly} against other functions?
Relative comparisons allow an LLM to make a ``you know it when you see it''
judgment call without losing any critical context to the quantized score.
Instead of ``score this function according to the rubric,'' the problem becomes
``re-order this list of functions according to their apparent relevance to the
advisory.''

This approach of relatively ranking lists of functions worked well for small
datasets, but broke down with large inputs. LLMs struggled to attend to large
amounts of data and regularly failed to return all of the functions in the
original input. To overcome this problem, we developed a method of randomly
sampling the function dataset to select small batches of functions at a time,
and measured how well each function tended to rank against the advisory within
its relative batch.
We identified the high performers by repeatedly sampling the data and averaging
the relative positions of each item across each of its randomly selected
batches; high performers would consistently land in positions averaging between
\#1--2 within a random batch of 10 items. We then filtered down the working
dataset to only include those high-ranking items, and performed ranking again.
This
iterative, progressive reduction of the dataset would eventually narrow down to
a single highest-ranked item placed at the top of the reassembled ranked
dataset.

We first demonstrated this technique at RVAsec
2024\footnote{\url{https://www.youtube.com/watch?v=IBuL1zY69tY\&t=1846s}},
showing it could successfully identify fixed vulnerable functions in large
firmware patches. Following this initial demonstration, we formalized the
approach in a blog post at Bishop
Fox\footnote{\url{https://bishopfox.com/blog/raink-llms-document-ranking}}, at
which point we recognized connections to existing work in information retrieval (IR)
and learning-to-rank literature. We presented improved results at DistrictCon
2025\footnote{\url{https://www.youtube.com/watch?v=FIYKlv48f6Y\&t=1051s}}, and
introduced algorithm enhancements (particularly, inflection-based convergence
detection) at Offensive AI Con
2025\footnote{\url{https://noperator.dev/posts/on-the-money}}.

This work represents an independent discovery of listwise ranking methodology
driven by applied vulnerability research constraints rather than theoretical
optimization in the information retrieval domain. The algorithm emerged from
the practical need to process datasets that exceeded typical LLM context
windows, while also maintaining result quality and operating within reasonable
cost and time budgets. Only after demonstrating empirical success did we
retroactively map our approach onto the established framework of listwise
ranking methods.

\subsection{Contributions}

This paper presents two main contributions:

\begin{itemize}

	\item We present \textbf{SiftRank}, a listwise document ranking algorithm
	      that achieves $O(n)$ complexity through stochastic sampling,
	      inflection-based convergence detection, and a fixed number of
	      iterative trials. This design enables consistent and efficient
	      ranking of large datasets.

	\item We demonstrate that complex security problems can be
	      \textit{transformed} into document ranking problems and subsequently
	      solved via information retrieval algorithms. For example, rather than
	      treating patch-diffing as a domain-specific problem requiring
	      specialized security engineering knowledge, we reframe it as ranking
	      changed functions (documents) by their relevance to a security
	      advisory (query).

\end{itemize}

\section{Related Work}

\subsection{High-Level Ranking Approaches}

Document ranking with LLMs can be approached in three ways: pointwise,
pairwise, and listwise.
\textbf{Pointwise} ranking assigns a numerical relevance score to one document 
at a time, and sorts the resulting scores to identify the document that is most 
relevant to the given query.
Documents are not compared directly against one another, but are rather reduced 
to a numerical score that can be easily sorted.
This approach achieves $O(n)$ complexity but yields inconsistent results as 
LLMs struggle to assign absolute scores to a single document in isolation.
In contrast, \textbf{pairwise} ranking performs an A/B comparison on 
\textit{two} documents at a time and makes a relative decision about which 
document is most relevant to the query.
This approach takes advantage of an LLM's ability to handle abstract tasks, and 
allows an LLM to easily substitute as a nondeterministic comparator in classic 
sorting algorithms like quicksort and heapsort.
Pairwise ranking is generally limited to $O(n^2)$ or $O(n \log n)$ performance.
\textbf{Listwise} ranking compares \textit{multiple} documents at once and 
performs a relative ordering of the dataset, similar to pairwise ranking. A 
listwise approach has potential for $O(n)$ performance, but faces considerable 
implementation challenges
clearly outlined by \cite{qin-etal-2024-large}, including failures where LLMs
output incomplete lists, refuse to rank, repeat items, or produce inconsistent
rankings across multiple executions. Some of these challenges are shared by
pointwise and pairwise rankers, but are exacerbated by listwise ranking's
larger operating dataset.

\subsection{Ranking with LLMs}

\cite{sun-etal-2023-chatgpt} introduced \textbf{RankGPT}, demonstrating that 
large language models could perform zero-shot listwise document ranking. Their 
approach partitions documents into fixed-size sliding windows, prompts the LLM 
to rank items within each window, then merges results by processing windows 
sequentially from the bottom of the list toward the top. This sliding window 
strategy processes all documents through deterministic windows, achieving 
$O(n)$ complexity with constant window size.
RankGPT operates exclusively as a reranker, requiring a separate first-stage
retriever to narrow the corpus to approximately 100 candidates before applying
LLM-based ranking. This means that its effectiveness is limited by first-stage
recall performance; if the initial retriever misses relevant documents, no
amount of sophisticated reranking can recover them.

\cite{Liu2023LostIT} reported the performance degradation that LLM rankers 
experience when changing the initial input order of documents. In response, 
\cite{tang-etal-2024-found} introduced \textbf{Permutation Self-Consistency}, 
which addresses this positional bias through stochastic sampling and 
aggregation. Their method repeatedly shuffles the input document list 
(typically 20 permutations), obtains a ranking from the LLM for each shuffle, 
then aggregates these rankings. This approach shares our fundamental insight 
that multiple stochastic trials can average out positional bias and LLM 
inconsistencies.
However, Permutation Self-Consistency faces a scalability constraint in that it
requires all documents to fit within the LLM's context window simultaneously.
For their sorting tasks, they rank only 10 items at once. For passage reranking
with 100 documents, they borrow RankGPT's sliding window approach and apply
their shuffle-and-aggregate method to each window separately, limiting their
approach to datasets of at most a few hundred documents.

\cite{10.1145/3626772.3657813} presented \textbf{Setwise}, which treats the LLM 
as a comparator within classic sorting algorithms. Rather than asking the LLM 
to rank pairs of documents, Setwise prompts it to identify the single most 
relevant document from a set of items (typically 3--10)
and performs top-$k$ extraction.
By using heapsort structure with these set-based comparisons, Setwise achieves
$O(\log n)$ complexity. However, like RankGPT, Setwise operates exclusively as
a reranker. Additionally, the deterministic heapsort structure provides no
mechanism for error resilience; a single incorrect judgment by an LLM
comparator can permanently eliminate a relevant document from top-$k$
consideration.

\cite{wang-etal-2025-realm} introduced \textbf{REALM}, a technique that 
performs setwise ranking on small groups of documents and extracts relevance 
probabilities from the LLM's output logits. It progressively narrows down the 
dataset by using high-confidence pivots to repeatedly split the working dataset 
until the algorithm converges on the most relevant documents. REALM shares our 
approach of reducing the dataset according to dynamically calculated relevance 
scores. Like other approaches above, it primarily operates as a reranker and 
relies on first-stage retrieval of relevant data. REALM also only supports 
providers and models that expose log probabilities of each potential output 
token.

\subsection{Ranking for Patch Identification}

\cite{Li2024PatchFinderAT} developed PatchFinder as a way to associate CVE 
descriptions to corresponding patches in open-source repositories. It operates 
as a two-stage information retrieval pipeline, with the first retrieval phase 
using keyword matching and code embeddings to narrow down commits to the most 
promising candidates, and the second reranking phase using a fine-tuned model 
to locate the actual patch. PatchFinder shares our approach of using IR 
techniques to match commits to a vulnerability description, but requires 
labeled training data and domain-specific model infrastructure rather than 
operating zero-shot with general-purpose LLMs. It also uses separate techniques 
for retrieval and reranking stages.

\section{The SiftRank Algorithm}

\textbf{SiftRank} 
(\textbf{S}tochastic-\textbf{I}nflective-\textbf{F}ixed-\textbf{T}rial Rank) is 
a listwise ranking algorithm that employs large language models as 
general-purpose document rankers for arbitrary datasets. We use the generic 
term ``document'' to mean the basic unit of retrieval in a large corpus of 
data. Depending on the dataset, a document could be a web page, source code 
snippet, JSON object, etc. The ranker does not assume any consistent schema 
among incoming data, so documents of varying or dissimilar types may be mixed 
and ranked against one another. SiftRank is designed to operate with $O(n)$ 
complexity so that it completes in a reasonable amount of time---fast enough to 
be used in a practitioner's real-time workflow. This algorithm is named for the 
qualities that allow it to \textit{sift} through very large datasets:
\begin{itemize}[nosep]
	\item \textbf{S}tochastic: Randomly samples small batches of documents to average out positional bias.
	\item \textbf{I}nflective: Detects emergent threshold between relevant and irrelevant documents.
	\item \textbf{F}ixed: Enforces a capped number of trials to ensure worst-case linear complexity.
	\item \textbf{T}rial: Repeatedly ranks sampled batches across iterations until convergence.
\end{itemize}
\bigskip
An open-source implementation of SiftRank exists at \url{https://github.com/noperator/siftrank}.

\subsection{Problem Formulation}
Given a corpus $C = \{d_1, d_2, \ldots, d_n\}$ of $n$ documents and a query $q$
defining the relevance criteria, the goal is to efficiently identify the
top-ranked documents in $C$ according to their relevance to $q$. We assume
access to an LLM ranking function $L(B, q)$ that can order a batch of documents
$B \subset C$ in descending order according to each document's relevance to
$q$. We also assume the ability to measure a dataset's inflection point $\tau$,
the point of maximum curvature in the sorted score distribution. The key
parameters are:
\begin{itemize}[nosep]
	\item Batch size $S$ (must fit within the LLM's context window, e.g.,
	      $S=10$)
	\item Maximum trials $T$ (fixed upper bound to ensure $O(n)$ complexity,
	      e.g., $T=50$)
	\item Stability window $W$ (number of consecutive trials required for
	      convergence, e.g., $W=5$)
\end{itemize}

\subsection{Algorithm Description}

SiftRank operates through multiple iterations $k = 1, 2, \ldots$ on
successively smaller subsets $C_k \subset C$ until $|C_k| = 1$. Initially, $C_1
= C$. For each iteration $k$ on the current corpus $C_k$:

\begin{enumerate}
	\item \textbf{Ranking:} For trials $t = 1, 2, \ldots, T$:
	      \begin{enumerate}
		      \item Randomly shuffle $C_k$ to obtain $C_k^{(t)}$, the working
		            dataset for trial $t$
		      \item Partition $C_k^{(t)}$ into $m = \lfloor |C_k^{(t)}|/S
		            \rfloor$ disjoint batches $B_1, B_2, \ldots, B_m$, each of
		            size $S$
		      \item For each batch $B_i$ where $i = 1, \ldots, m$, rank the
		            batch by applying $L(B_i, q)$ and record each document's
		            relative position within the ranked batch, $p_d^{(t)} \in
		            \{1, 2, \ldots, S\}$ (where the document at position 1 has
		            the highest relevance to the query $q$)
		      \item Update running average scores across completed trials:
		            $s_d^{(t)} = \frac{1}{t} \sum_{i=1}^{t} p_d^{(i)}$
		      \item Sort documents by score $s_d^{(t)}$ to obtain ranking
		            $R_k^{(t)}$, and measure inflection point $\tau_k^{(t)}$

		      \item \textbf{Convergence:} Determine if convergence trial
		      $t^*$ has been reached. For the past $W$ consecutively
		      completed trials:
		            \begin{enumerate}
			            \item If ordering stabilizes ($R_k^{(t)} = R_k^{(t-i)}$
			                  for $i = 1, \ldots, W-1$), set $t^* = t$ and $K =
			                  k$, and \textbf{go to Step 3}
			            \item If inflection point stabilizes ($\tau_k^{(t)} =
			                  \tau_k^{(t-i)}$ for $i = 1, \ldots, W-1$), set
			                  $t^* = t$ and \textbf{go to Step 2}
		            \end{enumerate}
		      \item If $t = T$ (maximum trials reached without convergence),
		            set $t^* = T$ and \textbf{go to Step 2}
	      \end{enumerate}

	      \newpage

	\item \textbf{Refinement:}
	      \begin{enumerate}
		      \item Partition the final ranking $R_k = R_k^{(t^*)}$ at
		            inflection point $\tau_k = \tau_k^{(t^*)}$:
		            \begin{itemize}[nosep]
			            \item $C_{k+1} = \{d \in R_k : s_d^{(t^*)} \leq \tau_k\}$ (top portion, advances to next iteration)

			            \item $F_k = \{d \in R_k : s_d^{(t^*)} > \tau_k\}$ (bottom portion, frozen in ordering from $R_k$)
		            \end{itemize}
		      \item If $|C_{k+1}| > 1$, \textbf{go to Step 1} with corpus
		            $C_{k+1}$ at iteration $k+1$
		      \item Otherwise, set $K = k$ and \textbf{go to Step 3}
	      \end{enumerate}

	\item \textbf{Reassembly:} With $K$ as the final iteration, return final ranked corpus by concatenating:
	      $$R = R_K, F_{K-1}, F_{K-2}, \ldots, F_2, F_1$$
\end{enumerate}

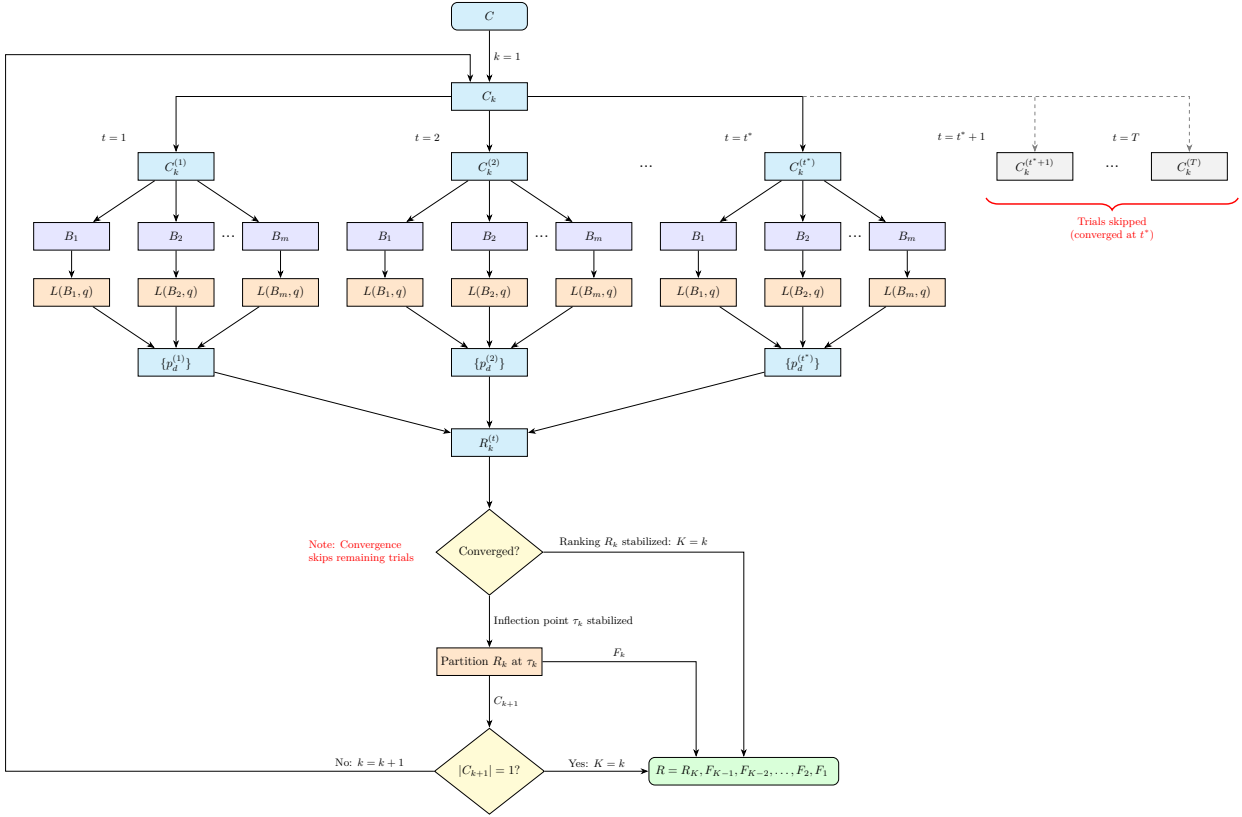
\begin{figure}[p]
	\centering

	\resizebox{\textwidth}{!}{
		\begin{tikzpicture}[
				node distance=1.5cm,
				box/.style={rectangle, draw, thick, minimum width=2.2cm, minimum height=0.8cm, align=center, fill=white},
				roundbox/.style={rectangle, rounded corners=5pt, draw, thick, minimum width=2.2cm, minimum height=0.8cm, align=center},
				trial/.style={rectangle, draw, thick, minimum width=2.2cm, minimum height=0.8cm, align=center, fill=cyan!15},
				batch/.style={rectangle, draw, thick, minimum width=2.2cm, minimum height=0.8cm, align=center, fill=blue!10},
				result/.style={rectangle, draw, thick, minimum width=2.2cm, minimum height=0.8cm, align=center, fill=green!15},
				decision/.style={diamond, draw, thick, minimum width=2.5cm, minimum height=2.5cm, align=center, fill=yellow!20, aspect=2},
				arrow/.style={-Stealth, thick},
				label/.style={font=\small}
			]

			\coordinate (dataset_center) at (0, 0);

			\node[batch, below=7cm of dataset_center, xshift=-10cm] (batch1_1) {$B_1$};
			\node[batch, right=0.8cm of batch1_1] (batch1_2) {$B_2$};
			\node[batch, right=0.8cm of batch1_2] (batch1_3) {$B_m$};
			\node[at=($(batch1_2)!0.5!(batch1_3)$), font=\Large] (batch1_dots) {...};

			\node[box, below=0.8cm of batch1_1, fill=orange!20] (llm1_1) {$L(B_1, q)$};
			\node[box, below=0.8cm of batch1_2, fill=orange!20] (llm1_2) {$L(B_2, q)$};
			\node[box, below=0.8cm of batch1_3, fill=orange!20] (llm1_3) {$L(B_m, q)$};

			\node[box, below=1.2cm of llm1_2, fill=cyan!15] (pos1) {$\{p_d^{(1)}\}$};

			\node[batch, right=0.8cm of batch1_3] (batch2_1) {$B_1$};
			\node[batch, right=0.8cm of batch2_1] (batch2_2) {$B_2$};
			\node[batch, right=0.8cm of batch2_2] (batch2_3) {$B_m$};
			\node[at=($(batch2_2)!0.5!(batch2_3)$), font=\Large] (batch2_dots) {...};

			\node[box, below=0.8cm of batch2_1, fill=orange!20] (llm2_1) {$L(B_1, q)$};
			\node[box, below=0.8cm of batch2_2, fill=orange!20] (llm2_2) {$L(B_2, q)$};
			\node[box, below=0.8cm of batch2_3, fill=orange!20] (llm2_3) {$L(B_m, q)$};

			\node[box, below=1.2cm of llm2_2, fill=cyan!15] (pos2) {$\{p_d^{(2)}\}$};

			\node[batch, right=0.8cm of batch2_3] (batchc_1) {$B_1$};
			\node[batch, right=0.8cm of batchc_1] (batchc_2) {$B_2$};
			\node[batch, right=0.8cm of batchc_2] (batchc_3) {$B_m$};
			\node[at=($(batchc_2)!0.5!(batchc_3)$), font=\Large] (batchc_dots) {...};

			\node[box, below=0.8cm of batchc_1, fill=orange!20] (llmc_1) {$L(B_1, q)$};
			\node[box, below=0.8cm of batchc_2, fill=orange!20] (llmc_2) {$L(B_2, q)$};
			\node[box, below=0.8cm of batchc_3, fill=orange!20] (llmc_3) {$L(B_m, q)$};

			\node[box, below=1.2cm of llmc_2, fill=cyan!15] (posc) {$\{p_d^{(t^*)}\}$};

			\node[trial, above=1.2cm of batch1_2] (trial1) {$C_k^{(1)}$};
			\node[trial, above=1.2cm of batch2_2] (trial2) {$C_k^{(2)}$};
			\node[trial, above=1.2cm of batchc_2] (trialc) {$C_k^{(t^*)}$};
			\node[at=($(trial2)!0.5!(trialc)$), font=\Large] (dots1) {...};
			\node[trial, right=4.5cm of trialc, fill=gray!10] (trialc1) {$C_k^{(t^*+1)}$};
			\node[trial, right=2.25cm of trialc1, fill=gray!10] (trialt) {$C_k^{(T)}$};
			\node[at=($(trialc1)!0.5!(trialt)$), font=\Large] (dots2) {...};

			\node[above left=0.2cm and 0.2cm of trial1, font=\small] {$t = 1$};
			\node[above left=0.2cm and 0.2cm of trial2, font=\small] {$t = 2$};
			\node[above left=0.2cm and 0.2cm of trialc, font=\small] {$t = t^*$};
			\node[above left=0.2cm and 0.2cm of trialc1, font=\small] {$t = t^*+1$};
			\node[above left=0.2cm and 0.2cm of trialt, font=\small] {$t = T$};

			\node[box, fill=cyan!15, above=1.2cm of trial2] (dataset) {$C_k$};

			\node[box, below=1.5cm of pos2, fill=cyan!15] (cumulative_rank) {$R_k^{(t)}$};

			\node[decision, below=1.5cm of cumulative_rank, minimum height=2.5cm] (converge) {Converged?};

			\node[box, below=1.5cm of converge, fill=orange!20] (partition) {Partition $R_k$ at $\tau_k$};

			\node[decision, below=1.5cm of partition] (size_check) {$|C_{k+1}| = 1$?};

			\node[roundbox, right=3cm of size_check, minimum width=5.5cm, align=center, fill=green!15] (final_result) {
				$R = R_K , F_{K-1} , F_{K-2} , \ldots , F_2 , F_1$
			};

			\draw[arrow] (dataset) -| (trial1);
			\draw[arrow] (dataset) -- (trial2);
			\coordinate (bend_point) at (trialc |- dataset);
			\draw[arrow] (dataset) -| (bend_point) -- (trialc);
			\draw[arrow, dashed, gray] (bend_point) -| (trialc1);
			\draw[arrow, dashed, gray] (bend_point) -| (trialt);

			\draw[arrow] (trial1) -- (batch1_1);
			\draw[arrow] (trial1) -- (batch1_2);
			\draw[arrow] (trial1) -- (batch1_3);

			\draw[arrow] (batch1_1) -- (llm1_1);
			\draw[arrow] (batch1_2) -- (llm1_2);
			\draw[arrow] (batch1_3) -- (llm1_3);

			\draw[arrow] (llm1_1) -- (pos1);
			\draw[arrow] (llm1_2) -- (pos1);
			\draw[arrow] (llm1_3) -- (pos1);

			\draw[arrow] (trial2) -- (batch2_1);
			\draw[arrow] (trial2) -- (batch2_2);
			\draw[arrow] (trial2) -- (batch2_3);

			\draw[arrow] (batch2_1) -- (llm2_1);
			\draw[arrow] (batch2_2) -- (llm2_2);
			\draw[arrow] (batch2_3) -- (llm2_3);

			\draw[arrow] (llm2_1) -- (pos2);
			\draw[arrow] (llm2_2) -- (pos2);
			\draw[arrow] (llm2_3) -- (pos2);

			\draw[arrow] (trialc) -- (batchc_1);
			\draw[arrow] (trialc) -- (batchc_2);
			\draw[arrow] (trialc) -- (batchc_3);

			\draw[arrow] (batchc_1) -- (llmc_1);
			\draw[arrow] (batchc_2) -- (llmc_2);
			\draw[arrow] (batchc_3) -- (llmc_3);

			\draw[arrow] (llmc_1) -- (posc);
			\draw[arrow] (llmc_2) -- (posc);
			\draw[arrow] (llmc_3) -- (posc);

			\draw[arrow] (pos1) -- (cumulative_rank);
			\draw[arrow] (pos2) -- (cumulative_rank);
			\draw[arrow] (posc) -- (cumulative_rank);

			\draw[arrow] (cumulative_rank) -- (converge);

			\coordinate (brace_left) at ($(trialc1.south west) + (-0.3, -0.5)$);
			\coordinate (brace_right) at ($(trialt.south east) + (0.3, -0.5)$);
			\coordinate (brace_center) at ($(brace_left)!0.5!(brace_right)$);
			\draw[decorate, decoration={brace, amplitude=10pt, mirror}, red, very thick]
			(brace_left |- brace_left) -- (brace_right |- brace_left)
			node[midway, below=12pt, font=\small, red, align=center] (skip_label) {Trials skipped\\(converged at $t^*$)};

			\node[label, left=0.5cm of converge, align=left, font=\small, red] {Note: Convergence\\skips remaining trials};

			\draw[arrow] (converge.south) -- node[midway, right, font=\small] {Inflection point $\tau_k$ stabilized} (partition);
			\draw[arrow] (partition) -- node[midway, right, font=\small] {$C_{k+1}$} (size_check);

			\coordinate (fk_entry) at ($(final_result.north west)!0.25!(final_result.north east)$);
			\coordinate (fk_turn) at (fk_entry |- partition.east);
			\draw[arrow] (partition.east) -- node[midway, above, font=\small] {$F_k$} (fk_turn) -- (fk_entry);

			\coordinate (converge_exit) at (converge.east);
			\coordinate (above_reassembly) at (converge_exit -| final_result.north);
			\draw[arrow] (converge_exit) -- node[pos=0.45, above, font=\small] {Ranking $R_k$ stabilized: $K=k$} (above_reassembly) -- (final_result.north);

			\draw[arrow] (size_check.east) -- (final_result.west) node[midway, above, font=\small] {Yes: $K=k$};

			\coordinate (loop_start) at (size_check.west);
			\coordinate (loop_far_left) at ($(batch1_1.west) + (-0.8, 0)$);
			\coordinate (loop_bottom) at (loop_far_left |- size_check);
			\coordinate (dataset_entry) at ($(dataset.north west)!0.25!(dataset.north east)$);
			\coordinate (loop_top_left) at ($(loop_far_left |- dataset.north) + (0, 0.8)$);
			\coordinate (loop_top_entry) at (loop_top_left -| dataset_entry);
			\draw[arrow, black] (loop_start) -- node[pos=0.15, above, font=\small] {No: $k = k+1$} (loop_bottom) -- (loop_top_left) -- (loop_top_entry) -- (dataset_entry);

			\node[roundbox, fill=cyan!15, above=1.5cm of dataset] (initial_corpus) {$C$};
			\draw[arrow, black] (initial_corpus) -- node[right, font=\small] {$k=1$} (dataset);

		\end{tikzpicture}
	}%
	\caption{SiftRank algorithm flow showing stochastic trial loop, batch partitioning, LLM ranking operations, convergence detection, and iterative refinement. The corpus $C_k$ is randomly shuffled for each trial $t$, partitioned into $m$ batches of size $S$, and ranked by the LLM in $L$. Positions $p$ are aggregated across trials to compute scores in $R_k^{(t)}$. When the inflection point $\tau_k$ stabilizes, the corpus is partitioned at that threshold, with top candidates $C_{k+1}$ advancing to the next iteration and frozen portions $F_k$ reserved for final reassembly.}
	\label{fig:siftrank-flow}
\end{figure}

Figure~\ref{fig:siftrank-flow} illustrates the complete algorithm flow, showing
how the corpus is shuffled into trials, batched for LLM ranking, and
iteratively refined until convergence.

\subsection{Implementation Considerations}
\label{sec:impl}

Several details of the algorithm are intentionally left unspecified and may
vary by implementation.

\begin{itemize}[nosep]

	\item Ranking concurrency: The trial loop is embarrassingly parallel. All
	      batches within all trials within a single iteration may be ranked
	      concurrently.

	\item Batch exclusion: Remainder documents excluded from batches in the
	      first trial (due to the floor operation) should be included in the
	      second trial, ensuring all documents are evaluated.

	\item LLM output: Each batch of documents should be presented to the
	      ranking model as a dictionary, and the ranking model should only
	      return the ordered keys. This significantly reduces the output token
	      generation requirements, saving time and inference cost.

	\item Score calculation: The median may also be used instead of mean, since
	      this makes the running score less sensitive to outliers.

	\item Convergence detection: Rather than requiring exact stable ordering or
	      inflection point values, we may instead accept variation within some
	      tolerance range.

	\item Inflection measurement: The inflection point $\tau$ may be measured
	      via elbow detection (identifying the point of maximum curvature in
	      the score distribution). Alternatively, simple gap detection (finding
	      the largest gap between consecutive scores) may be simpler to
	      implement but provides less reliable inflection signal.

	\item Iterative refinement: This process is more easily illustrated
	      iteratively, but may instead be completed recursively (see
	      aforementioned open-source implementation as an example).

	\item Document summarization: When individual documents are large enough to
	      strain context window constraints (particularly with smaller models),
	      an optional preprocessing step may distill each document into a
	      summary (optionally focused on its relevance to the query). The
	      algorithm then ranks these summaries rather than the full documents.
	      This additional step falls outside the core algorithm but proves
	      useful when working with context-constrained models or very large
	      documents.

	\item Relevance reasoning: The ranking model may optionally accompany the
	      ordered documents with an \textit{explanation} of its reasoning for
	      each batch ordering. This useful property helps a practitioner
	      provide feedback to the model by adjusting the query (prompt) to
	      steer the model's reasoning for future ranking attempts.

\end{itemize}

\newpage

\subsection{Illustrative Example}

To clearly illustrate how SiftRank works, we demonstrate the algorithm on a
simple concrete example of ranking top-level domain names (TLDs) by their
relevance to the concept of mathematics. This toy problem requires no domain
expertise to understand, yet exhibits the key algorithmic properties that
enable SiftRank to scale to complex security use cases.

\subsubsection{Problem Setup}

Consider a scenario where a student would like to register a personal domain to
express their enthusiasm for mathematics. Ideally, the TLD would relate to math
as closely as possible. Given 536 TLDs sampled from the IANA registry (e.g.,
\texttt{.com}, \texttt{.biz}, etc.), the goal is to identify the TLD that ranks
highest in response to the query, ``Which of these top-level domains relates
most closely to the concept of theoretical mathematics?''

\subsubsection{Algorithm Execution}

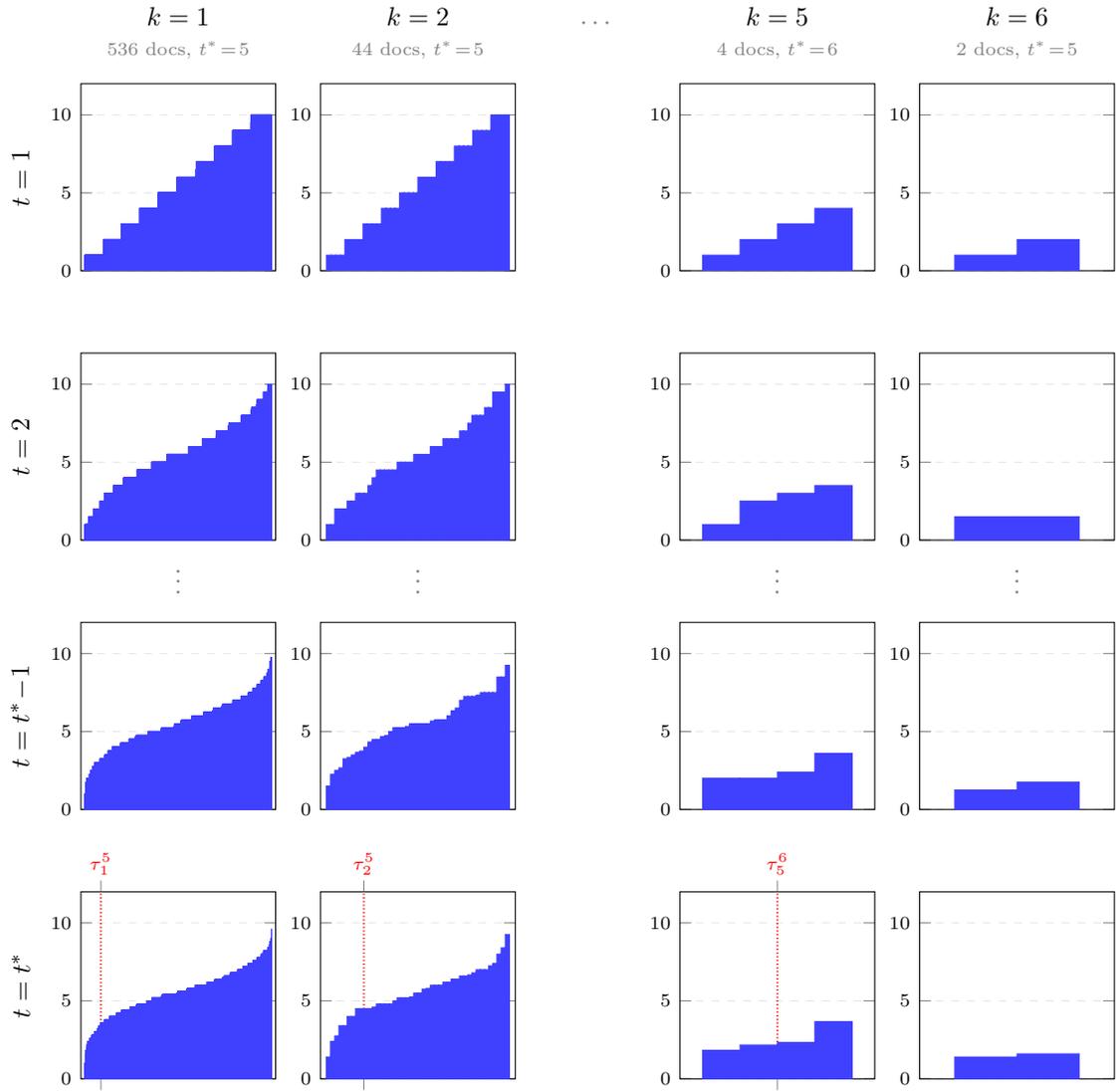
\begin{figure}[p]
	\centering

	\pgfplotsset{
		siftrank bar/.style={
				width=2.6cm,
				height=2.5cm,
				scale only axis,
				ybar,
				ymin=0,
				ymax=12,
				xmin=0,
				xtick=\empty,
				ytick={0,5,10},
				yticklabels={0,5,10},
				grid=major,
				grid style={dashed,gray!20},
				enlarge x limits=0.02,
				tick label style={font=\scriptsize},
				every axis plot post/.append style={fill=blue!75, draw=blue!75},
			}
	}

	\begin{tikzpicture}

		\node[font=\normalsize\bfseries] at (1.3, 0.6) {$k=1$};
		\node[font=\normalsize\bfseries] at (4.5, 0.6) {$k=2$};
		\node[font=\normalsize\bfseries] at (9.3, 0.6) {$k=5$};
		\node[font=\normalsize\bfseries] at (12.5, 0.6) {$k=6$};

		\node[font=\scriptsize, gray] at (1.3, 0.15) {536 docs, $t^*\!=\!5$};
		\node[font=\scriptsize, gray] at (4.5, 0.15) {44 docs, $t^*\!=\!5$};
		\node[font=\scriptsize, gray] at (9.3, 0.15) {4 docs, $t^*\!=\!6$};
		\node[font=\scriptsize, gray] at (12.5, 0.15) {2 docs, $t^*\!=\!5$};

		\node[font=\normalsize\bfseries, rotate=90, anchor=center] at (-0.8, -1.55) {$t=1$};
		\node[font=\normalsize\bfseries, rotate=90, anchor=center] at (-0.8, -5.15) {$t=2$};
		\node[font=\normalsize\bfseries, rotate=90, anchor=center] at (-0.8, -8.75) {$t=t^*\!-\!1$};
		\node[font=\normalsize\bfseries, rotate=90, anchor=center] at (-0.8, -12.35) {$t=t^*$};

		\node[font=\normalsize, gray] at (6.9, 0.5) {$\cdots$};

		\node[font=\normalsize, gray] at (1.3, -6.85) {$\vdots$};
		\node[font=\normalsize, gray] at (4.5, -6.85) {$\vdots$};
		\node[font=\normalsize, gray] at (9.3, -6.85) {$\vdots$};
		\node[font=\normalsize, gray] at (12.5, -6.85) {$\vdots$};

		\begin{axis}[siftrank bar, at={(0,-0.3cm)}, anchor=north west, xmax=531, bar width=0.14pt]
			\addplot coordinates {
					(1,1)(2,1)(3,1)(4,1)(5,1)(6,1)(7,1)(8,1)(9,1)(10,1)(11,1)(12,1)(13,1)(14,1)(15,1)(16,1)(17,1)(18,1)(19,1)(20,1)(21,1)(22,1)(23,1)(24,1)(25,1)(26,1)(27,1)(28,1)(29,1)(30,1)(31,1)(32,1)(33,1)(34,1)(35,1)(36,1)(37,1)(38,1)(39,1)(40,1)(41,1)(42,1)(43,1)(44,1)(45,1)(46,1)(47,1)(48,1)(49,1)(50,1)(51,1)(52,1)(53,2)(54,2)(55,2)(56,2)(57,2)(58,2)(59,2)(60,2)(61,2)(62,2)(63,2)(64,2)(65,2)(66,2)(67,2)(68,2)(69,2)(70,2)(71,2)(72,2)(73,2)(74,2)(75,2)(76,2)(77,2)(78,2)(79,2)(80,2)(81,2)(82,2)(83,2)(84,2)(85,2)(86,2)(87,2)(88,2)(89,2)(90,2)(91,2)(92,2)(93,2)(94,2)(95,2)(96,2)(97,2)(98,2)(99,2)(100,2)(101,2)(102,2)(103,2)(104,3)(105,3)(106,3)(107,3)(108,3)(109,3)(110,3)(111,3)(112,3)(113,3)(114,3)(115,3)(116,3)(117,3)(118,3)(119,3)(120,3)(121,3)(122,3)(123,3)(124,3)(125,3)(126,3)(127,3)(128,3)(129,3)(130,3)(131,3)(132,3)(133,3)(134,3)(135,3)(136,3)(137,3)(138,3)(139,3)(140,3)(141,3)(142,3)(143,3)(144,3)(145,3)(146,3)(147,3)(148,3)(149,3)(150,3)(151,3)(152,3)(153,3)(154,3)(155,3)(156,4)(157,4)(158,4)(159,4)(160,4)(161,4)(162,4)(163,4)(164,4)(165,4)(166,4)(167,4)(168,4)(169,4)(170,4)(171,4)(172,4)(173,4)(174,4)(175,4)(176,4)(177,4)(178,4)(179,4)(180,4)(181,4)(182,4)(183,4)(184,4)(185,4)(186,4)(187,4)(188,4)(189,4)(190,4)(191,4)(192,4)(193,4)(194,4)(195,4)(196,4)(197,4)(198,4)(199,4)(200,4)(201,4)(202,4)(203,4)(204,4)(205,4)(206,4)(207,4)(208,5)(209,5)(210,5)(211,5)(212,5)(213,5)(214,5)(215,5)(216,5)(217,5)(218,5)(219,5)(220,5)(221,5)(222,5)(223,5)(224,5)(225,5)(226,5)(227,5)(228,5)(229,5)(230,5)(231,5)(232,5)(233,5)(234,5)(235,5)(236,5)(237,5)(238,5)(239,5)(240,5)(241,5)(242,5)(243,5)(244,5)(245,5)(246,5)(247,5)(248,5)(249,5)(250,5)(251,5)(252,5)(253,5)(254,5)(255,5)(256,5)(257,5)(258,5)(259,5)(260,5)(261,5)(262,6)(263,6)(264,6)(265,6)(266,6)(267,6)(268,6)(269,6)(270,6)(271,6)(272,6)(273,6)(274,6)(275,6)(276,6)(277,6)(278,6)(279,6)(280,6)(281,6)(282,6)(283,6)(284,6)(285,6)(286,6)(287,6)(288,6)(289,6)(290,6)(291,6)(292,6)(293,6)(294,6)(295,6)(296,6)(297,6)(298,6)(299,6)(300,6)(301,6)(302,6)(303,6)(304,6)(305,6)(306,6)(307,6)(308,6)(309,6)(310,6)(311,6)(312,6)(313,6)(314,6)(315,6)(316,6.5)(317,7)(318,7)(319,7)(320,7)(321,7)(322,7)(323,7)(324,7)(325,7)(326,7)(327,7)(328,7)(329,7)(330,7)(331,7)(332,7)(333,7)(334,7)(335,7)(336,7)(337,7)(338,7)(339,7)(340,7)(341,7)(342,7)(343,7)(344,7)(345,7)(346,7)(347,7)(348,7)(349,7)(350,7)(351,7)(352,7)(353,7)(354,7)(355,7)(356,7)(357,7)(358,7)(359,7)(360,7)(361,7)(362,7)(363,7)(364,7)(365,7)(366,7)(367,7)(368,7)(369,8)(370,8)(371,8)(372,8)(373,8)(374,8)(375,8)(376,8)(377,8)(378,8)(379,8)(380,8)(381,8)(382,8)(383,8)(384,8)(385,8)(386,8)(387,8)(388,8)(389,8)(390,8)(391,8)(392,8)(393,8)(394,8)(395,8)(396,8)(397,8)(398,8)(399,8)(400,8)(401,8)(402,8)(403,8)(404,8)(405,8)(406,8)(407,8)(408,8)(409,8)(410,8)(411,8)(412,8)(413,8)(414,8)(415,8)(416,8)(417,8)(418,8)(419,8)(420,9)(421,9)(422,9)(423,9)(424,9)(425,9)(426,9)(427,9)(428,9)(429,9)(430,9)(431,9)(432,9)(433,9)(434,9)(435,9)(436,9)(437,9)(438,9)(439,9)(440,9)(441,9)(442,9)(443,9)(444,9)(445,9)(446,9)(447,9)(448,9)(449,9)(450,9)(451,9)(452,9)(453,9)(454,9)(455,9)(456,9)(457,9)(458,9)(459,9)(460,9)(461,9)(462,9)(463,9)(464,9)(465,9)(466,9)(467,9)(468,9)(469,9)(470,9)(471,9.5)(472,10)(473,10)(474,10)(475,10)(476,10)(477,10)(478,10)(479,10)(480,10)(481,10)(482,10)(483,10)(484,10)(485,10)(486,10)(487,10)(488,10)(489,10)(490,10)(491,10)(492,10)(493,10)(494,10)(495,10)(496,10)(497,10)(498,10)(499,10)(500,10)(501,10)(502,10)(503,10)(504,10)(505,10)(506,10)(507,10)(508,10)(509,10)(510,10)(511,10)(512,10)(513,10)(514,10)(515,10)(516,10)(517,10)(518,10)(519,10)(520,10)(521,10)(522,10)(523,10)(524,10)(525,10)(526,10)(527,10)(528,10)(529,10)(530,10)
				};
		\end{axis}

		\begin{axis}[siftrank bar, at={(3.2cm,-0.3cm)}, anchor=north west, xmax=41, bar width=1.77pt]
			\addplot coordinates {
					(1,1)(2,1)(3,1)(4,1)(5,2)(6,2)(7,2)(8,2)(9,3)(10,3)(11,3)(12,3)(13,4)(14,4)(15,4)(16,4)(17,5)(18,5)(19,5)(20,5)(21,6)(22,6)(23,6)(24,6)(25,7)(26,7)(27,7)(28,7)(29,8)(30,8)(31,8)(32,8)(33,9)(34,9)(35,9)(36,9)(37,10)(38,10)(39,10)(40,10)
				};
		\end{axis}

		\begin{axis}[siftrank bar, at={(8.0cm,-0.3cm)}, anchor=north west, xmax=5, bar width=14.15pt]
			\addplot coordinates {(1,1)(2,2)(3,3)(4,4)};
		\end{axis}

		\begin{axis}[siftrank bar, at={(11.2cm,-0.3cm)}, anchor=north west, xmax=3, bar width=23.6pt]
			\addplot coordinates {(1,1)(2,2)};
		\end{axis}

		\begin{axis}[siftrank bar, at={(0,-3.9cm)}, anchor=north west, xmax=537, bar width=0.13pt]
			\addplot coordinates {
					(1,1)(2,1)(3,1)(4,1)(5,1)(6,1)(7,1)(8,1)(9,1)(10,1)(11,1.5)(12,1.5)(13,1.5)(14,1.5)(15,1.5)(16,1.5)(17,1.5)(18,1.5)(19,1.5)(20,1.5)(21,1.5)(22,1.5)(23,1.5)(24,1.5)(25,2)(26,2)(27,2)(28,2)(29,2)(30,2)(31,2)(32,2)(33,2)(34,2)(35,2)(36,2)(37,2)(38,2)(39,2)(40,2)(41,2)(42,2)(43,2)(44,2.5)(45,2.5)(46,2.5)(47,2.5)(48,2.5)(49,2.5)(50,2.5)(51,2.5)(52,2.5)(53,2.5)(54,2.5)(55,2.5)(56,2.5)(57,3)(58,3)(59,3)(60,3)(61,3)(62,3)(63,3)(64,3)(65,3)(66,3)(67,3)(68,3)(69,3)(70,3)(71,3)(72,3)(73,3)(74,3)(75,3)(76,3)(77,3)(78,3)(79,3)(80,3)(81,3)(82,3)(83,3.5)(84,3.5)(85,3.5)(86,3.5)(87,3.5)(88,3.5)(89,3.5)(90,3.5)(91,3.5)(92,3.5)(93,3.5)(94,3.5)(95,3.5)(96,3.5)(97,3.5)(98,3.5)(99,3.5)(100,3.5)(101,3.5)(102,3.5)(103,3.5)(104,3.5)(105,3.5)(106,3.5)(107,3.5)(108,3.5)(109,3.5)(110,3.5)(111,3.5)(112,4)(113,4)(114,4)(115,4)(116,4)(117,4)(118,4)(119,4)(120,4)(121,4)(122,4)(123,4)(124,4)(125,4)(126,4)(127,4)(128,4)(129,4)(130,4)(131,4)(132,4)(133,4)(134,4)(135,4)(136,4)(137,4)(138,4)(139,4)(140,4)(141,4)(142,4)(143,4)(144,4)(145,4)(146,4)(147,4)(148,4)(149,4)(150,4.5)(151,4.5)(152,4.5)(153,4.5)(154,4.5)(155,4.5)(156,4.5)(157,4.5)(158,4.5)(159,4.5)(160,4.5)(161,4.5)(162,4.5)(163,4.5)(164,4.5)(165,4.5)(166,4.5)(167,4.5)(168,4.5)(169,4.5)(170,4.5)(171,4.5)(172,4.5)(173,4.5)(174,4.5)(175,4.5)(176,4.5)(177,4.5)(178,4.5)(179,4.5)(180,4.5)(181,4.5)(182,4.5)(183,4.5)(184,4.5)(185,4.5)(186,4.5)(187,4.5)(188,4.5)(189,4.5)(190,4.5)(191,4.5)(192,5)(193,5)(194,5)(195,5)(196,5)(197,5)(198,5)(199,5)(200,5)(201,5)(202,5)(203,5)(204,5)(205,5)(206,5)(207,5)(208,5)(209,5)(210,5)(211,5)(212,5)(213,5)(214,5)(215,5)(216,5)(217,5)(218,5)(219,5)(220,5)(221,5)(222,5)(223,5)(224,5)(225,5)(226,5)(227,5)(228,5)(229,5)(230,5)(231,5)(232,5)(233,5)(234,5)(235,5)(236,5.5)(237,5.5)(238,5.5)(239,5.5)(240,5.5)(241,5.5)(242,5.5)(243,5.5)(244,5.5)(245,5.5)(246,5.5)(247,5.5)(248,5.5)(249,5.5)(250,5.5)(251,5.5)(252,5.5)(253,5.5)(254,5.5)(255,5.5)(256,5.5)(257,5.5)(258,5.5)(259,5.5)(260,5.5)(261,5.5)(262,5.5)(263,5.5)(264,5.5)(265,5.5)(266,5.5)(267,5.5)(268,5.5)(269,5.5)(270,5.5)(271,5.5)(272,5.5)(273,5.5)(274,5.5)(275,5.5)(276,5.5)(277,5.5)(278,5.5)(279,5.5)(280,5.5)(281,5.5)(282,5.5)(283,5.5)(284,5.5)(285,5.5)(286,5.5)(287,5.5)(288,5.5)(289,5.5)(290,5.5)(291,5.5)(292,5.5)(293,5.5)(294,5.5)(295,5.5)(296,5.5)(297,5.5)(298,6)(299,6)(300,6)(301,6)(302,6)(303,6)(304,6)(305,6)(306,6)(307,6)(308,6)(309,6)(310,6)(311,6)(312,6)(313,6)(314,6)(315,6)(316,6)(317,6)(318,6)(319,6)(320,6)(321,6)(322,6)(323,6)(324,6)(325,6)(326,6)(327,6)(328,6)(329,6)(330,6)(331,6)(332,6)(333,6)(334,6)(335,6)(336,6)(337,6)(338,6.5)(339,6.5)(340,6.5)(341,6.5)(342,6.5)(343,6.5)(344,6.5)(345,6.5)(346,6.5)(347,6.5)(348,6.5)(349,6.5)(350,6.5)(351,6.5)(352,6.5)(353,6.5)(354,6.5)(355,6.5)(356,6.5)(357,6.5)(358,6.5)(359,6.5)(360,6.5)(361,6.5)(362,6.5)(363,6.5)(364,6.5)(365,6.5)(366,6.5)(367,6.5)(368,6.5)(369,6.5)(370,6.5)(371,6.5)(372,6.5)(373,6.5)(374,6.5)(375,6.5)(376,6.5)(377,6.67)(378,7)(379,7)(380,7)(381,7)(382,7)(383,7)(384,7)(385,7)(386,7)(387,7)(388,7)(389,7)(390,7)(391,7)(392,7)(393,7)(394,7)(395,7)(396,7)(397,7)(398,7)(399,7)(400,7)(401,7)(402,7)(403,7)(404,7)(405,7)(406,7)(407,7)(408,7)(409,7)(410,7)(411,7)(412,7)(413,7.33)(414,7.5)(415,7.5)(416,7.5)(417,7.5)(418,7.5)(419,7.5)(420,7.5)(421,7.5)(422,7.5)(423,7.5)(424,7.5)(425,7.5)(426,7.5)(427,7.5)(428,7.5)(429,7.5)(430,7.5)(431,7.5)(432,7.5)(433,7.5)(434,7.5)(435,7.5)(436,7.5)(437,7.5)(438,7.5)(439,7.5)(440,7.5)(441,7.5)(442,7.5)(443,7.5)(444,7.5)(445,7.5)(446,7.5)(447,7.5)(448,7.5)(449,7.67)(450,8)(451,8)(452,8)(453,8)(454,8)(455,8)(456,8)(457,8)(458,8)(459,8)(460,8)(461,8)(462,8)(463,8)(464,8)(465,8)(466,8)(467,8)(468,8)(469,8)(470,8)(471,8)(472,8)(473,8)(474,8)(475,8)(476,8)(477,8)(478,8.33)(479,8.5)(480,8.5)(481,8.5)(482,8.5)(483,8.5)(484,8.5)(485,8.5)(486,8.5)(487,8.5)(488,8.5)(489,8.5)(490,8.5)(491,8.5)(492,8.5)(493,8.67)(494,9)(495,9)(496,9)(497,9)(498,9)(499,9)(500,9)(501,9)(502,9)(503,9)(504,9)(505,9)(506,9)(507,9)(508,9)(509,9)(510,9)(511,9)(512,9)(513,9.5)(514,9.5)(515,9.5)(516,9.5)(517,9.5)(518,9.5)(519,9.5)(520,9.5)(521,9.5)(522,9.5)(523,9.5)(524,9.5)(525,9.67)(526,10)(527,10)(528,10)(529,10)(530,10)(531,10)(532,10)(533,10)(534,10)(535,10)(536,10)
				};
		\end{axis}

		\begin{axis}[siftrank bar, at={(3.2cm,-3.9cm)}, anchor=north west, xmax=45, bar width=1.6pt]
			\addplot coordinates {
					(1,1)(2,1)(3,2)(4,2)(5,2)(6,2.5)(7,2.5)(8,3)(9,3)(10,3)(11,3.5)(12,4)(13,4.5)(14,4.5)(15,4.5)(16,4.5)(17,4.5)(18,5)(19,5)(20,5)(21,5)(22,5.5)(23,5.5)(24,5.5)(25,5.5)(26,6)(27,6)(28,6)(29,6.5)(30,6.5)(31,6.5)(32,6.5)(33,7)(34,7)(35,7.5)(36,8)(37,8)(38,8)(39,8.5)(40,8.5)(41,9.5)(42,9.5)(43,9.5)(44,10)
				};
		\end{axis}

		\begin{axis}[siftrank bar, at={(8.0cm,-3.9cm)}, anchor=north west, xmax=5, bar width=14.15pt]
			\addplot coordinates {(1,1)(2,2.5)(3,3)(4,3.5)};
		\end{axis}

		\begin{axis}[siftrank bar, at={(11.2cm,-3.9cm)}, anchor=north west, xmax=3, bar width=23.6pt]
			\addplot coordinates {(1,1.5)(2,1.5)};
		\end{axis}

		\begin{axis}[siftrank bar, at={(0,-7.5cm)}, anchor=north west, xmax=537, bar width=0.13pt]
			\addplot coordinates {
					(1,1)(2,1)(3,1.5)(4,1.75)(5,1.75)(6,2)(7,2)(8,2)(9,2)(10,2)(11,2)(12,2)(13,2.25)(14,2.25)(15,2.25)(16,2.25)(17,2.5)(18,2.5)(19,2.5)(20,2.5)(21,2.5)(22,2.5)(23,2.75)(24,2.75)(25,2.75)(26,2.75)(27,2.75)(28,2.75)(29,3)(30,3)(31,3)(32,3)(33,3)(34,3)(35,3)(36,3)(37,3)(38,3)(39,3)(40,3)(41,3)(42,3)(43,3)(44,3.25)(45,3.25)(46,3.25)(47,3.25)(48,3.25)(49,3.25)(50,3.25)(51,3.25)(52,3.25)(53,3.25)(54,3.25)(55,3.25)(56,3.25)(57,3.25)(58,3.5)(59,3.5)(60,3.5)(61,3.5)(62,3.5)(63,3.5)(64,3.5)(65,3.5)(66,3.5)(67,3.75)(68,3.75)(69,3.75)(70,3.75)(71,3.75)(72,3.75)(73,3.75)(74,3.75)(75,3.75)(76,3.75)(77,3.75)(78,3.75)(79,4)(80,4)(81,4)(82,4)(83,4)(84,4)(85,4)(86,4)(87,4)(88,4)(89,4)(90,4)(91,4)(92,4)(93,4)(94,4)(95,4)(96,4)(97,4)(98,4)(99,4)(100,4)(101,4)(102,4)(103,4.25)(104,4.25)(105,4.25)(106,4.25)(107,4.25)(108,4.25)(109,4.25)(110,4.25)(111,4.25)(112,4.25)(113,4.25)(114,4.25)(115,4.25)(116,4.25)(117,4.25)(118,4.25)(119,4.25)(120,4.25)(121,4.25)(122,4.25)(123,4.25)(124,4.25)(125,4.25)(126,4.25)(127,4.5)(128,4.5)(129,4.5)(130,4.5)(131,4.5)(132,4.5)(133,4.5)(134,4.5)(135,4.5)(136,4.5)(137,4.5)(138,4.5)(139,4.5)(140,4.5)(141,4.5)(142,4.5)(143,4.5)(144,4.5)(145,4.6)(146,4.67)(147,4.67)(148,4.75)(149,4.75)(150,4.75)(151,4.75)(152,4.75)(153,4.75)(154,4.75)(155,4.75)(156,4.75)(157,4.75)(158,4.75)(159,4.75)(160,4.75)(161,4.75)(162,4.75)(163,4.75)(164,4.75)(165,4.75)(166,4.75)(167,4.75)(168,4.75)(169,4.75)(170,4.75)(171,4.75)(172,4.75)(173,4.75)(174,4.75)(175,4.75)(176,4.75)(177,4.75)(178,4.75)(179,4.75)(180,4.75)(181,4.75)(182,5)(183,5)(184,5)(185,5)(186,5)(187,5)(188,5)(189,5)(190,5)(191,5)(192,5)(193,5)(194,5)(195,5)(196,5)(197,5)(198,5)(199,5)(200,5)(201,5)(202,5)(203,5)(204,5)(205,5)(206,5)(207,5)(208,5)(209,5)(210,5)(211,5)(212,5)(213,5)(214,5)(215,5)(216,5)(217,5)(218,5)(219,5)(220,5.2)(221,5.25)(222,5.25)(223,5.25)(224,5.25)(225,5.25)(226,5.25)(227,5.25)(228,5.25)(229,5.25)(230,5.25)(231,5.25)(232,5.25)(233,5.25)(234,5.25)(235,5.25)(236,5.25)(237,5.25)(238,5.25)(239,5.25)(240,5.25)(241,5.25)(242,5.25)(243,5.25)(244,5.25)(245,5.25)(246,5.25)(247,5.25)(248,5.25)(249,5.25)(250,5.25)(251,5.25)(252,5.25)(253,5.25)(254,5.25)(255,5.25)(256,5.33)(257,5.33)(258,5.5)(259,5.5)(260,5.5)(261,5.5)(262,5.5)(263,5.5)(264,5.5)(265,5.5)(266,5.5)(267,5.5)(268,5.5)(269,5.5)(270,5.5)(271,5.5)(272,5.5)(273,5.5)(274,5.5)(275,5.5)(276,5.67)(277,5.67)(278,5.67)(279,5.75)(280,5.75)(281,5.75)(282,5.75)(283,5.75)(284,5.75)(285,5.75)(286,5.75)(287,5.75)(288,5.75)(289,5.75)(290,5.75)(291,5.75)(292,5.75)(293,5.75)(294,5.75)(295,5.75)(296,5.75)(297,5.75)(298,5.75)(299,5.75)(300,5.75)(301,5.75)(302,5.75)(303,5.75)(304,5.75)(305,5.75)(306,5.75)(307,5.8)(308,6)(309,6)(310,6)(311,6)(312,6)(313,6)(314,6)(315,6)(316,6)(317,6)(318,6)(319,6)(320,6)(321,6)(322,6)(323,6)(324,6)(325,6)(326,6)(327,6)(328,6)(329,6)(330,6)(331,6)(332,6)(333,6)(334,6)(335,6)(336,6)(337,6)(338,6)(339,6)(340,6)(341,6.2)(342,6.25)(343,6.25)(344,6.25)(345,6.25)(346,6.25)(347,6.25)(348,6.25)(349,6.25)(350,6.25)(351,6.25)(352,6.25)(353,6.25)(354,6.25)(355,6.25)(356,6.25)(357,6.25)(358,6.25)(359,6.25)(360,6.25)(361,6.25)(362,6.25)(363,6.25)(364,6.25)(365,6.25)(366,6.25)(367,6.33)(368,6.33)(369,6.4)(370,6.5)(371,6.5)(372,6.5)(373,6.5)(374,6.5)(375,6.5)(376,6.5)(377,6.5)(378,6.5)(379,6.5)(380,6.5)(381,6.5)(382,6.5)(383,6.5)(384,6.5)(385,6.5)(386,6.5)(387,6.5)(388,6.5)(389,6.5)(390,6.5)(391,6.5)(392,6.5)(393,6.6)(394,6.6)(395,6.67)(396,6.75)(397,6.75)(398,6.75)(399,6.75)(400,6.75)(401,6.75)(402,6.75)(403,6.75)(404,6.75)(405,6.75)(406,6.75)(407,6.75)(408,6.75)(409,6.75)(410,6.75)(411,6.75)(412,6.75)(413,6.75)(414,6.75)(415,6.75)(416,6.75)(417,6.75)(418,6.75)(419,6.75)(420,6.75)(421,6.75)(422,6.75)(423,6.75)(424,6.8)(425,7)(426,7)(427,7)(428,7)(429,7)(430,7)(431,7)(432,7)(433,7)(434,7)(435,7)(436,7)(437,7)(438,7)(439,7)(440,7)(441,7)(442,7)(443,7)(444,7)(445,7)(446,7)(447,7)(448,7)(449,7.25)(450,7.25)(451,7.25)(452,7.25)(453,7.25)(454,7.25)(455,7.25)(456,7.25)(457,7.25)(458,7.25)(459,7.25)(460,7.25)(461,7.25)(462,7.25)(463,7.25)(464,7.25)(465,7.25)(466,7.25)(467,7.25)(468,7.25)(469,7.33)(470,7.5)(471,7.5)(472,7.5)(473,7.5)(474,7.5)(475,7.5)(476,7.5)(477,7.5)(478,7.5)(479,7.5)(480,7.5)(481,7.5)(482,7.75)(483,7.75)(484,7.75)(485,7.75)(486,7.75)(487,7.75)(488,7.75)(489,7.75)(490,7.75)(491,7.75)(492,7.8)(493,7.8)(494,8)(495,8)(496,8)(497,8)(498,8)(499,8)(500,8)(501,8)(502,8)(503,8)(504,8)(505,8.25)(506,8.25)(507,8.25)(508,8.25)(509,8.25)(510,8.25)(511,8.25)(512,8.25)(513,8.25)(514,8.5)(515,8.5)(516,8.5)(517,8.5)(518,8.5)(519,8.5)(520,8.5)(521,8.5)(522,8.5)(523,8.75)(524,8.75)(525,8.75)(526,8.75)(527,9)(528,9)(529,9)(530,9)(531,9)(532,9.5)(533,9.5)(534,9.5)(535,9.75)(536,9.75)
				};
		\end{axis}

		\begin{axis}[siftrank bar, at={(3.2cm,-7.5cm)}, anchor=north west, xmax=45, bar width=1.6pt]
			\addplot coordinates {
					(1,1.5)(2,2.25)(3,2.5)(4,2.67)(5,3.25)(6,3.33)(7,3.5)(8,3.67)(9,3.75)(10,4)(11,4.33)(12,4.5)(13,4.5)(14,4.67)(15,4.75)(16,5)(17,5.25)(18,5.25)(19,5.25)(20,5.33)(21,5.5)(22,5.5)(23,5.5)(24,5.5)(25,5.5)(26,5.67)(27,5.75)(28,5.75)(29,5.75)(30,6)(31,6.33)(32,6.5)(33,7)(34,7.25)(35,7.25)(36,7.25)(37,7.33)(38,7.5)(39,7.5)(40,7.5)(41,7.5)(42,8.5)(43,8.5)(44,9.25)
				};
		\end{axis}

		\begin{axis}[siftrank bar, at={(8.0cm,-7.5cm)}, anchor=north west, xmax=5, bar width=14.15pt]
			\addplot coordinates {(1,2)(2,2)(3,2.4)(4,3.6)};
		\end{axis}

		\begin{axis}[siftrank bar, at={(11.2cm,-7.5cm)}, anchor=north west, xmax=3, bar width=23.6pt]
			\addplot coordinates {(1,1.25)(2,1.75)};
		\end{axis}

		\begin{axis}[siftrank bar, at={(0,-11.1cm)}, anchor=north west, xmax=537, bar width=0.13pt, clip=false,
				extra x ticks={46.5}, extra x tick style={grid=major, grid style={red, densely dotted, line width=0.8pt}, xticklabel={\empty}}]
			\addplot coordinates {
					(1,1)(2,1.6)(3,1.8)(4,1.8)(5,1.8)(6,2)(7,2.2)(8,2.4)(9,2.4)(10,2.4)(11,2.4)(12,2.4)(13,2.4)(14,2.5)(15,2.6)(16,2.6)(17,2.6)(18,2.6)(19,2.6)(20,2.6)(21,2.8)(22,2.8)(23,2.8)(24,2.8)(25,2.8)(26,2.8)(27,2.8)(28,3)(29,3)(30,3)(31,3)(32,3)(33,3)(34,3)(35,3)(36,3.2)(37,3.2)(38,3.2)(39,3.2)(40,3.25)(41,3.4)(42,3.4)(43,3.4)(44,3.4)(45,3.4)(46,3.6)(47,3.6)(48,3.6)(49,3.6)(50,3.6)(51,3.6)(52,3.6)(53,3.6)(54,3.6)(55,3.6)(56,3.75)(57,3.75)(58,3.8)(59,3.8)(60,3.8)(61,3.8)(62,3.8)(63,3.8)(64,3.8)(65,3.8)(66,3.8)(67,3.8)(68,3.8)(69,3.8)(70,3.8)(71,4)(72,4)(73,4)(74,4)(75,4)(76,4)(77,4)(78,4)(79,4)(80,4)(81,4)(82,4)(83,4)(84,4)(85,4)(86,4)(87,4)(88,4)(89,4)(90,4)(91,4.2)(92,4.2)(93,4.2)(94,4.2)(95,4.2)(96,4.2)(97,4.2)(98,4.2)(99,4.2)(100,4.2)(101,4.2)(102,4.2)(103,4.2)(104,4.25)(105,4.33)(106,4.4)(107,4.4)(108,4.4)(109,4.4)(110,4.4)(111,4.4)(112,4.4)(113,4.4)(114,4.4)(115,4.4)(116,4.4)(117,4.4)(118,4.4)(119,4.4)(120,4.4)(121,4.4)(122,4.4)(123,4.4)(124,4.4)(125,4.4)(126,4.4)(127,4.4)(128,4.4)(129,4.4)(130,4.5)(131,4.6)(132,4.6)(133,4.6)(134,4.6)(135,4.6)(136,4.6)(137,4.6)(138,4.6)(139,4.6)(140,4.6)(141,4.6)(142,4.6)(143,4.6)(144,4.6)(145,4.6)(146,4.6)(147,4.6)(148,4.6)(149,4.75)(150,4.75)(151,4.8)(152,4.8)(153,4.8)(154,4.8)(155,4.8)(156,4.8)(157,4.8)(158,4.8)(159,4.8)(160,4.8)(161,4.8)(162,4.8)(163,4.8)(164,4.8)(165,4.8)(166,4.8)(167,4.8)(168,4.8)(169,4.8)(170,4.8)(171,4.8)(172,4.8)(173,4.8)(174,4.8)(175,4.8)(176,4.8)(177,4.8)(178,5)(179,5)(180,5)(181,5)(182,5)(183,5)(184,5)(185,5)(186,5)(187,5)(188,5)(189,5)(190,5)(191,5)(192,5)(193,5.2)(194,5.2)(195,5.2)(196,5.2)(197,5.2)(198,5.2)(199,5.2)(200,5.2)(201,5.2)(202,5.2)(203,5.2)(204,5.2)(205,5.2)(206,5.2)(207,5.2)(208,5.2)(209,5.2)(210,5.2)(211,5.2)(212,5.2)(213,5.2)(214,5.2)(215,5.2)(216,5.2)(217,5.2)(218,5.2)(219,5.25)(220,5.25)(221,5.33)(222,5.33)(223,5.4)(224,5.4)(225,5.4)(226,5.4)(227,5.4)(228,5.4)(229,5.4)(230,5.4)(231,5.4)(232,5.4)(233,5.4)(234,5.4)(235,5.4)(236,5.4)(237,5.4)(238,5.4)(239,5.4)(240,5.4)(241,5.4)(242,5.4)(243,5.4)(244,5.4)(245,5.4)(246,5.4)(247,5.4)(248,5.4)(249,5.4)(250,5.4)(251,5.4)(252,5.4)(253,5.4)(254,5.4)(255,5.4)(256,5.4)(257,5.4)(258,5.4)(259,5.4)(260,5.4)(261,5.4)(262,5.4)(263,5.4)(264,5.4)(265,5.5)(266,5.5)(267,5.6)(268,5.6)(269,5.6)(270,5.6)(271,5.6)(272,5.6)(273,5.6)(274,5.6)(275,5.6)(276,5.6)(277,5.6)(278,5.6)(279,5.6)(280,5.6)(281,5.6)(282,5.6)(283,5.6)(284,5.6)(285,5.6)(286,5.6)(287,5.6)(288,5.6)(289,5.6)(290,5.6)(291,5.6)(292,5.6)(293,5.75)(294,5.75)(295,5.8)(296,5.8)(297,5.8)(298,5.8)(299,5.8)(300,5.8)(301,5.8)(302,5.8)(303,5.8)(304,5.8)(305,5.8)(306,5.8)(307,5.8)(308,5.8)(309,5.8)(310,5.8)(311,5.8)(312,5.8)(313,5.8)(314,5.8)(315,5.83)(316,5.83)(317,6)(318,6)(319,6)(320,6)(321,6)(322,6)(323,6)(324,6)(325,6)(326,6)(327,6)(328,6)(329,6)(330,6)(331,6)(332,6)(333,6)(334,6)(335,6)(336,6)(337,6)(338,6)(339,6)(340,6)(341,6)(342,6)(343,6)(344,6)(345,6)(346,6)(347,6)(348,6)(349,6)(350,6)(351,6.2)(352,6.2)(353,6.2)(354,6.2)(355,6.2)(356,6.2)(357,6.2)(358,6.2)(359,6.2)(360,6.2)(361,6.2)(362,6.2)(363,6.2)(364,6.2)(365,6.2)(366,6.2)(367,6.2)(368,6.2)(369,6.2)(370,6.2)(371,6.2)(372,6.25)(373,6.33)(374,6.4)(375,6.4)(376,6.4)(377,6.4)(378,6.4)(379,6.4)(380,6.4)(381,6.4)(382,6.4)(383,6.4)(384,6.4)(385,6.4)(386,6.4)(387,6.4)(388,6.4)(389,6.4)(390,6.4)(391,6.4)(392,6.4)(393,6.4)(394,6.4)(395,6.4)(396,6.5)(397,6.5)(398,6.5)(399,6.5)(400,6.6)(401,6.6)(402,6.6)(403,6.6)(404,6.6)(405,6.6)(406,6.6)(407,6.6)(408,6.6)(409,6.6)(410,6.6)(411,6.6)(412,6.6)(413,6.6)(414,6.6)(415,6.6)(416,6.6)(417,6.75)(418,6.8)(419,6.8)(420,6.8)(421,6.8)(422,6.8)(423,6.8)(424,6.8)(425,6.8)(426,6.8)(427,6.8)(428,6.8)(429,6.8)(430,6.8)(431,6.8)(432,6.8)(433,6.8)(434,6.8)(435,6.8)(436,6.8)(437,6.8)(438,6.8)(439,7)(440,7)(441,7)(442,7)(443,7)(444,7)(445,7)(446,7)(447,7)(448,7)(449,7)(450,7)(451,7)(452,7)(453,7)(454,7.2)(455,7.2)(456,7.2)(457,7.2)(458,7.2)(459,7.2)(460,7.2)(461,7.2)(462,7.2)(463,7.2)(464,7.2)(465,7.2)(466,7.2)(467,7.2)(468,7.25)(469,7.4)(470,7.4)(471,7.4)(472,7.4)(473,7.4)(474,7.4)(475,7.4)(476,7.4)(477,7.4)(478,7.4)(479,7.4)(480,7.4)(481,7.4)(482,7.4)(483,7.4)(484,7.6)(485,7.6)(486,7.6)(487,7.6)(488,7.6)(489,7.6)(490,7.6)(491,7.6)(492,7.6)(493,7.6)(494,7.6)(495,7.75)(496,7.8)(497,7.8)(498,7.8)(499,7.8)(500,7.8)(501,7.8)(502,7.8)(503,7.8)(504,7.8)(505,7.8)(506,7.83)(507,8)(508,8)(509,8)(510,8)(511,8)(512,8)(513,8.2)(514,8.2)(515,8.2)(516,8.2)(517,8.2)(518,8.2)(519,8.2)(520,8.2)(521,8.2)(522,8.2)(523,8.2)(524,8.25)(525,8.4)(526,8.4)(527,8.4)(528,8.4)(529,8.4)(530,8.6)(531,8.8)(532,8.8)(533,8.8)(534,8.8)(535,9)(536,9.6)
				};
			\node[red, font=\scriptsize, anchor=south] at (axis cs:46.5,12.5) {$\tau_1^{5}$};
		\end{axis}

		\begin{axis}[siftrank bar, at={(3.2cm,-11.1cm)}, anchor=north west, xmax=45, bar width=1.6pt, clip=false,
				extra x ticks={9.5}, extra x tick style={grid=major, grid style={red, densely dotted, line width=0.8pt}, xticklabel={\empty}}]
			\addplot coordinates {
					(1,1.4)(2,2.4)(3,2.75)(4,3.4)(5,3.4)(6,4)(7,4)(8,4.5)(9,4.5)(10,4.5)(11,4.5)(12,4.6)(13,4.8)(14,4.8)(15,4.8)(16,4.8)(17,5)(18,5.2)(19,5.2)(20,5.2)(21,5.25)(22,5.5)(23,5.5)(24,5.75)(25,5.8)(26,6)(27,6)(28,6)(29,6.2)(30,6.2)(31,6.4)(32,6.4)(33,6.6)(34,6.6)(35,6.67)(36,6.8)(37,7)(38,7)(39,7)(40,7.25)(41,7.4)(42,8)(43,8.4)(44,9.25)
				};
			\node[red, font=\scriptsize, anchor=south] at (axis cs:9.5,12.5) {$\tau_2^{5}$};
		\end{axis}

		\begin{axis}[siftrank bar, at={(8.0cm,-11.1cm)}, anchor=north west, xmax=5, bar width=14.15pt, clip=false,
				extra x ticks={2.5}, extra x tick style={grid=major, grid style={red, densely dotted, line width=0.8pt}, xticklabel={\empty}}]
			\addplot coordinates {(1,1.83)(2,2.17)(3,2.33)(4,3.67)};
			\node[red, font=\scriptsize, anchor=south] at (axis cs:2.5,12.5) {$\tau_5^{6}$};
		\end{axis}

		\begin{axis}[siftrank bar, at={(11.2cm,-11.1cm)}, anchor=north west, xmax=3, bar width=23.6pt, clip=false]
			\addplot coordinates {(1,1.4)(2,1.6)};
		\end{axis}

	\end{tikzpicture}

	\caption{Progressive emergence of inflection point in TLD score distributions.  \\
		\textit{Rows}: Trials 1, 2,~\ldots, $t^*\!-\!1$, $t^*$, where $t^*$ is the trial where the position of the inflection point stabilized.
		\textit{Columns}: Iterations 1, 2,~\ldots, $K\!-\!1$, $K$ where $K\!=\!6$.
		The red dotted line marks the position of the inflection point $\tau$ at convergence.}
	\label{fig:score-distributions}
\end{figure}

We configure SiftRank with batch size $S = 10$, maximum trials $T = 50$, and
stability window $W = 5$, and walk through the algorithm's execution following
the same steps outlined in its description:%

\begin{itemize}

	\item \textbf{Trial 1:} The 536 TLDs are randomly shuffled and partitioned into 53 batches of 10 items each.
	      Each batch is ranked independently by the LLM. For example, one batch might contain
	      \{~\texttt{.travel},
	      \texttt{.careers},
	      \texttt{.university},
	      \texttt{.education},
	      \texttt{.one},
	      \texttt{.gratis},
	      \texttt{.show},
	      \texttt{.academy},
	      \texttt{.viajes},
	      \texttt{.dance}~\}, which the LLM ranks as
	      \{~\#1~\texttt{.academy},
	      \#2~\texttt{.university},
	      \#3~\texttt{.education},
	      \#4~\texttt{.one},
	      \#5~\texttt{.travel},
	      \#6~\texttt{.careers},
	      \#7~\texttt{.gratis},
	      \#8~\texttt{.viajes},
	      \#9~\texttt{.dance},
	      \#10~\texttt{.show}~\}.
	      Each TLD receives its relative position
	      within its batch as an initial score, where a score closer to 1
	      indicates a higher degree of relevance.

	\item \textbf{Trials 2--5:} The dataset is reshuffled and batched for
	each trial, making sure to include the 6 remainders from trial 1.
	      Across trials, \texttt{.science}, \texttt{.academy}, and
	      \texttt{.degree} consistently receive low (highly relevant) position
	      scores regardless of which other TLDs appear in their batches. Their
	      average scores decrease and stabilize, while less relevant TLDs like
	      \texttt{.pizza} or \texttt{.ventures} accumulate higher average
	      position scores.

	\item \textbf{Convergence Detection:} After 5 trials, we observe the
	      score distribution. Highly relevant
	      items cluster at low scores (around 1--3), while less relevant
	      items spread across higher scores ($\ge$ 5). The inflection point
	      $\tau_5$ emerges at approximately score 3.6, identifying a natural
	      separation threshold where the rate of score increase changes
	      sharply.  See Figure~\ref{fig:score-distributions} for a
	      visualization of convergence detection.

	\item \textbf{Iterative Refinement:} TLDs with scores below $\tau_5$
	      (approximately 44 items) form
	      $C_2$ for iterative refinement. These top candidates undergo
	      additional trials with reshuffling to determine their relative
	      ordering. The algorithm continues through 6 rounds until a single
	      top-ranked item emerges.

\end{itemize}

\subsubsection{Results}

The final ranking places \texttt{.phd}, \texttt{.science}, and \texttt{.degree}
as the top three results---intuitively relevant answers that align with casual
human judgment about math-related domains. The entire process completes in 12
seconds at a negligible cost of \$0.04 using OpenAI's
\texttt{gpt-5-nano-2025-08-07}\footnote{Priced per million tokens at \$0.05
(input) and \$0.4 (output) at the time of this writing.} with \texttt{minimal}
reasoning effort.

\newpage

\section{Security Application}
To demonstrate SiftRank's practical effectiveness, we applied it to a
real-world N-day vulnerability analysis problem of identifying which functions
in a vendor firmware patch were actually responsible for fixing a disclosed
vulnerability.

\subsection{Problem Setup}
In January 2025, SonicWall disclosed
CVE-2024-53704,
an authentication bypass vulnerability in their SonicOS firewall firmware. The
security
advisory\footnote{\url{https://www.zerodayinitiative.com/advisories/ZDI-25-012/}}
described the vulnerability in general terms (``authentication bypass,''
``processing of Base64-encoded session cookies,'' ``incorrect implementation of
an authentication algorithm'') but did not specify the exact location of the
fix in the codebase. Security researchers performing N-day analysis must locate
the vulnerable function by examining the firmware patch, a task that becomes
increasingly difficult as patch size grows.

The SonicWall patch consisted of 2,197 changed functions in a stripped binary,
meaning it removed function names and debug symbols that would normally assist
with reverse-engineering and analysis. Manually examining this many functions
would require days of analyst time and deep domain expertise. We transformed
this into a ranking problem: given the security advisory text as a query and
decompiled code as documents, rank the functions by their relevance to fixing
the described vulnerability.

\subsection{Methodology}
We diffed the binary with BinDiff to identify which functions changed in the
patch. Unlike typical patch analysis workflows that narrow down changed
functions by BinDiff's similarity and confidence scores, we included
\textit{all} 2,197 changed functions without arbitrary thresholding, allowing
the ranking process to naturally handle noise and peripheral changes. We then
used Binary Ninja to extract and decompile the \textit{original} unpatched
definitions\footnote{The relatively smaller function code diff could be used
here rather than the entire function definition, but that loses rich context
about what the function is actually doing.} for each changed function. We used
an LLM to generate a brief summary of each function with the following prompt: ``In just a few
sentences, summarize what this function appears to be doing. Provide roughly 3
sentences of medium-level technical explanation (e.g., if a developer were
speaking to a technical product manager), and then 1 sentence of high-level
business explanation (e.g., if a technical product manager were speaking to a
sales representative).''

Examining individual functions in isolation provides limited signal about their
role in a potential vulnerability. To capture crucial interprocedural context,
we constructed a call graph from the binary patch and generated call chains of
length 1--2 (an individual function A, or a function pair where B calls C).
This expanded our dataset from 2,197 individual functions to 2,713 call chains,
providing larger but richer contextual information about how changed functions
actually interact.

We then ranked these call chains using SiftRank (batch size $S = 5$, maximum
trials $T = 50$, stability window $W = 5$), treating the CVE advisory text as
the ranking query. Following a retrieval-inspired approach, we discarded all
items that were eliminated in the first ranking iteration, keeping only the 254
call chains that survived multiple ranking iterations ($k > 1$), demonstrating
measurable relevance to the vulnerability description.

\newpage

\subsection{Cluster Analysis}
To identify the actual vulnerability location from the ranked call chains, we
performed a cluster analysis on the call graph. First, we extracted the
function name(s) from each ranked chain and assigned each function $f$ a weight
using the formula: $$w_f = \frac{k_f}{r_f}$$ where $r_f$ is the function's best
(lowest) rank across all call chains containing it, and $k_f$ is the maximum
number of ranking iterations the function survived. This weighting captures
both quality (lower rank indicates higher relevance) and confidence (higher
iteration count indicates greater stability).

Next, we reconstructed the ranked functions into clusters (i.e., subgraphs of
the greater call graph) where the cluster size is constrained by its diameter
(the maximum distance of one node to another). We created clusters using
diameter values of 1--3, pooled all resulting clusters, and then ranked
clusters by mass $\times$ density, where mass is the sum of function weights in
the cluster $C$ and density is the average weight: $$\text{score} = \text{mass}
\times \text{density} = \left(\sum_{f \in C} w_f\right) \times
\frac{1}{|C|}\left(\sum_{f \in C} w_f\right) = \frac{\left(\sum_{f \in C}
w_f\right)^2}{|C|}$$

\begin{figure}[t]
	\centering

	\begin{tikzpicture}[
			node distance=2cm,
			func/.style={rectangle, draw, thick, rounded corners=3pt, minimum width=5cm, minimum height=1.4cm, align=center, font=\footnotesize},
			rank1/.style={func, fill=cyan!50},
			rank2/.style={func, fill=cyan!35},
			rank3/.style={func, fill=cyan!22},
			rank4/.style={func, fill=cyan!12},
			arrow/.style={-Stealth, thick, gray!70}
		]

		\node[rank2] (ab01f0) at (0, 0) {\texttt{sub\_2ab01f0} {\scriptsize (rank=2, w=2.5)}\\[2pt]{\scriptsize Crafts auth reply to VPN client}\\[-1pt]{\scriptsize (success, next-steps, or error)}};
		\node[rank1] (aac220) at (7, 0) {\texttt{sub\_2aac220} {\scriptsize (rank=1, w=5.0)}\\[2pt]{\scriptsize Validates request session token and}\\[-1pt]{\scriptsize hands off payload for processing}};

		\node[rank1] (acb160) at (3.5, -2.25) {\texttt{sub\_2acb160} {\scriptsize (rank=1, w=5.0)}\\[2pt]{\scriptsize Extracts cookie-based identifier}\\[-1pt]{\scriptsize and retrieves SSL-VPN session}};

		\node[rank3] (cbae10) at (0, -4.5) {\texttt{sub\_2cbae10} {\scriptsize (rank=3, w=1.33)}\\[2pt]{\scriptsize Dispatcher that filters invalid}\\[-1pt]{\scriptsize inputs before logging}};
		\node[rank4] (acc210) at (7, -4.5) {\texttt{sub\_2acc210} {\scriptsize (rank=4, w=1.0)}\\[2pt]{\scriptsize Validates token exists}\\[-1pt]{\scriptsize in session cache}};

		\draw[arrow] (aac220) -- (acb160);
		\draw[arrow] (ab01f0) -- (acb160);
		\draw[arrow] (ab01f0) -- (cbae10);
		\draw[arrow] (acb160) -- (acc210);
		\draw[arrow] (acb160) -- (cbae10);

	\end{tikzpicture}

	\caption{Out of 2,713 function call chains (which were then grouped into 119 function call clusters), this weighted cluster ranked at the top. It clearly shows relevance to the security advisory which mentioned
		``authentication'' and ``session cookies,'' each of which are mentioned in the function summaries. We are able to surface the critically relevant (but relatively lower-weight) session validation function \texttt{sub\_2acc210} because of its association with other higher-weight functions in the cluster.}
	\label{fig:cluster}
\end{figure}
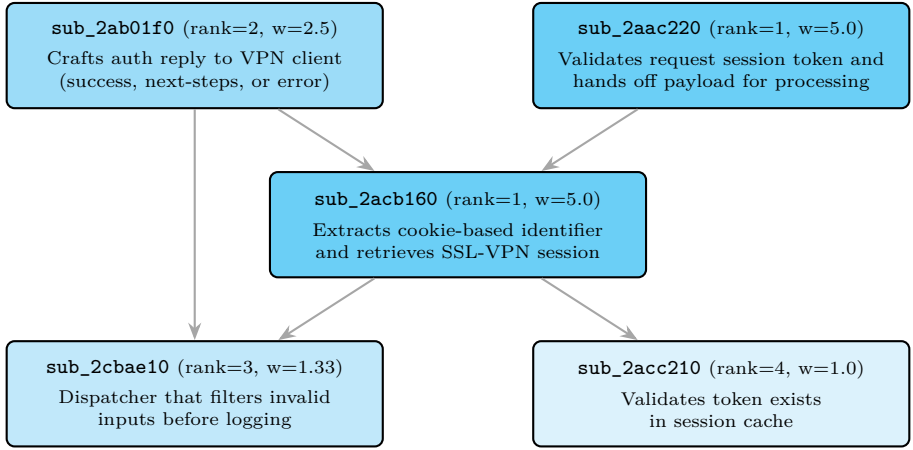

This mass-density metric naturally favors clusters that have multiple
high-ranking functions (mass) and also maintain concentrated relevance
(density).
The analysis identified 119 clusters across all diameter thresholds.
The top-ranked cluster (see Figure~\ref{fig:cluster}) contained 5 functions at
diameter 2, including the critical session validation function
\texttt{sub\_2acc210} which had a vulnerable string comparison loop that would
exit early and return a valid session when it encountered a null byte in an
attacker-supplied cookie.
\texttt{sub\_2cbae10} appears to be a logging function which likely ranked
highly simply because of its widespread use in the patched codebase. The other 3
functions in the cluster show clear relevance to authentication, and by
association help direct attention to the vulnerable token validation logic in
\texttt{sub\_2acc210}.
All 5 functions ranked within the top 4 call chains and demonstrated high
survival (4--5 iterations).
Manual verification against public vulnerability research\footnote{\url{https://bishopfox.com/blog/sonicwall-cve-2024-53704-ssl-vpn-session-hijacking}} confirmed
that these functions implement the Base64 cookie authentication mechanism
described in the CVE advisory, successfully localizing the vulnerability to
0.2\% of the patched codebase. See Appendix~\ref{app:chains} and
Appendix~\ref{app:clusters} for detailed rankings of call chains and clusters,
respectively.

\subsection{Results}
SiftRank efficiently identified the vulnerability-fixing functions in the
top-ranked cluster:
\smallskip
\begin{itemize}[nosep]
	\item \textbf{Accuracy:} Correct vulnerability cluster ranked \#1 of 119 clusters

	\item \textbf{Execution time\footnote{Summarizing took 65 seconds, and clustering took 21 seconds. Total operation took 3.08 minutes.}:} 99 seconds (specifically for SiftRank)
	\item \textbf{Inference cost\footnote{Total inference cost including summarization was \$1.47.}:} \$0.82 (using OpenAI's \texttt{gpt-5-nano-2025-08-07}, \texttt{minimal} reasoning)
	\item \textbf{LLM API calls:} 7,622 requests
	\item \textbf{Input tokens:} 14,007,343
	\item \textbf{Output tokens:} 303,378

\end{itemize}

\section{Discussion}

\subsection{Generality}

While we discovered and applied SiftRank primarily in security contexts, the
approach is fundamentally domain-agnostic. The algorithm makes no assumptions
about data structure or content---it requires only that an LLM can assess
relative relevance to a query among a small handful of items. This generality
enables application to any ranking problem where a context-sensitive judgment
call can distinguish relevant items from irrelevant ones.

SiftRank particularly excels at ``needle in a haystack'' problems where query
relevance is recognizable (``you know it when you see it'') but is not easily
quantifiable ahead of time. The LLM serves as a proxy for human intuitive
judgment at scale. This fuzzy specification capability extends beyond
security to domains like content discovery, research literature triage, or any
scenario requiring prioritization based on criteria that are easier to evaluate
than to specify. The approach works equally well on text, code, structured
data, or any content that an LLM can process. The algorithm treats all items as
opaque documents to be compared.

\subsection{Future Work}

This paper primarily serves the purpose of formalizing the SiftRank algorithm,
demonstrating its effectiveness on the use cases that motivated its
development, and attempting to connect the algorithm to related work in the
information retrieval domain. While the current evaluation focuses on a single
N-day analysis task, we have successfully used SiftRank to discover 0-day logic
vulnerabilities in widely used COTS appliances (e.g., using a query like,
``Which of these functions most likely contains an authentication bypass?'').
Future revisions of this paper may benchmark SiftRank's performance against
standard IR datasets including BEIR~\citep{Thakur2021Beir} and TREC Deep
Learning~\citep{Craswell2020TrecDl,Bajaj2016Msmarco} to enable direct
comparison with existing reranking methods.
We may also compare performance across closed- and open-weight models, expand
and improve the function clustering technique, and illustrate wider security
applications including source code analysis, web application testing, etc.

\section*{Acknowledgements}

Special thanks to Josh Shomo and Jon Williams for their early collaboration
toward solving security problems via document ranking with LLMs; to Emil
Gurevitch for inspiration around convergence detection; and to Trampas Howe,
Justin Rhinehart, and Daniel Cuthbert for reviewing early drafts of this paper.

\newpage

\bibliographystyle{plainnat}

\bibliography{references}

@inproceedings{sun-etal-2023-chatgpt,
    title = "Is {C}hat{GPT} Good at Search? Investigating Large Language Models as Re-Ranking Agents",
    author = "Sun, Weiwei  and
      Yan, Lingyong  and
      Ma, Xinyu  and
      Wang, Shuaiqiang  and
      Ren, Pengjie  and
      Chen, Zhumin  and
      Yin, Dawei  and
      Ren, Zhaochun",
    editor = "Bouamor, Houda  and
      Pino, Juan  and
      Bali, Kalika",
    booktitle = "Proceedings of the 2023 Conference on Empirical Methods in Natural Language Processing",
    month = dec,
    year = "2023",
    address = "Singapore",
    publisher = "Association for Computational Linguistics",
    url = "https://aclanthology.org/2023.emnlp-main.923/",
    doi = "10.18653/v1/2023.emnlp-main.923",
    pages = "14918--14937"
}

@inproceedings{tang-etal-2024-found,
    title = "Found in the Middle: Permutation Self-Consistency Improves Listwise Ranking in Large Language Models",
    author = "Tang, Raphael  and
      Zhang, Crystina  and
      Ma, Xueguang  and
      Lin, Jimmy  and
      Ture, Ferhan",
    editor = "Duh, Kevin  and
      Gomez, Helena  and
      Bethard, Steven",
    booktitle = "Proceedings of the 2024 Conference of the North American Chapter of the Association for Computational Linguistics: Human Language Technologies (Volume 1: Long Papers)",
    month = jun,
    year = "2024",
    address = "Mexico City, Mexico",
    publisher = "Association for Computational Linguistics",
    url = "https://aclanthology.org/2024.naacl-long.129/",
    doi = "10.18653/v1/2024.naacl-long.129",
    pages = "2327--2340"
}

@inproceedings{qin-etal-2024-large,
    title = "Large Language Models are Effective Text Rankers with Pairwise Ranking Prompting",
    author = "Qin, Zhen  and
      Jagerman, Rolf  and
      Hui, Kai  and
      Zhuang, Honglei  and
      Wu, Junru  and
      Yan, Le  and
      Shen, Jiaming  and
      Liu, Tianqi  and
      Liu, Jialu  and
      Metzler, Donald  and
      Wang, Xuanhui  and
      Bendersky, Michael",
    editor = "Duh, Kevin  and
      Gomez, Helena  and
      Bethard, Steven",
    booktitle = "Findings of the Association for Computational Linguistics: NAACL 2024",
    month = jun,
    year = "2024",
    address = "Mexico City, Mexico",
    publisher = "Association for Computational Linguistics",
    url = "https://aclanthology.org/2024.findings-naacl.97/",
    doi = "10.18653/v1/2024.findings-naacl.97",
    pages = "1504--1518"
}

@inproceedings{10.1145/3626772.3657813,
author = {Zhuang, Shengyao and Zhuang, Honglei and Koopman, Bevan and Zuccon, Guido},
title = {A Setwise Approach for Effective and Highly Efficient Zero-shot Ranking with Large Language Models},
year = {2024},
isbn = {9798400704314},
publisher = {Association for Computing Machinery},
address = {New York, NY, USA},
url = {https://doi.org/10.1145/3626772.3657813},
doi = {10.1145/3626772.3657813},
booktitle = {Proceedings of the 47th International ACM SIGIR Conference on Research and Development in Information Retrieval},
pages = {38–47},
numpages = {10},
location = {Washington DC, USA},
series = {SIGIR '24}
}

@article{Thakur2021Beir,
  title = "BEIR: A Heterogenous Benchmark for Zero-shot Evaluation of Information Retrieval Models",
  author = "Thakur, Nandan and Reimers, Nils and Rücklé, Andreas and Srivastava, Abhishek and Gurevych, Iryna", 
  journal= "arXiv preprint arXiv:2104.08663",
  month = "4",
  year = "2021",
  url = "https://arxiv.org/abs/2104.08663",
}

@inproceedings{Craswell2020TrecDl,
  title={Overview of the TREC 2020 deep learning track},
  author={Nick Craswell and Bhaskar Mitra and Emine Yilmaz and Daniel Campos},
  booktitle={TREC},
  year={2020}
}

@inproceedings{Bajaj2016Msmarco,
  title={MS MARCO: A Human Generated MAchine Reading COmprehension Dataset},
  author={Payal Bajaj and Daniel Campos and Nick Craswell and Li Deng and Jianfeng Gao and Xiaodong Liu and Rangan Majumder and Andrew McNamara and Bhaskar Mitra and Tri Nguyen and Mir Rosenberg and Xia Song and Alina Stoica and Saurabh Tiwary and Tong Wang},
  booktitle={InCoCo@NIPS},
  year={2016}
}

@article{Liu2023LostIT,
  title={Lost in the Middle: How Language Models Use Long Contexts},
  author={Nelson F. Liu and Kevin Lin and John Hewitt and Ashwin Paranjape and Michele Bevilacqua and Fabio Petroni and Percy Liang},
  journal={Transactions of the Association for Computational Linguistics},
  year={2023},
  volume={12},
  pages={157-173},
  url={https://api.semanticscholar.org/CorpusID:259360665}
}

@inproceedings{wang-etal-2025-realm,
    title = "{REALM}: Recursive Relevance Modeling for {LLM}-based Document Re-Ranking",
    author = "Wang, Pinhuan  and
      Xia, Zhiqiu  and
      Liao, Chunhua  and
      Wang, Feiyi  and
      Liu, Hang",
    editor = "Christodoulopoulos, Christos  and
      Chakraborty, Tanmoy  and
      Rose, Carolyn  and
      Peng, Violet",
    booktitle = "Proceedings of the 2025 Conference on Empirical Methods in Natural Language Processing",
    month = nov,
    year = "2025",
    address = "Suzhou, China",
    publisher = "Association for Computational Linguistics",
    url = "https://aclanthology.org/2025.emnlp-main.1218/",
    doi = "10.18653/v1/2025.emnlp-main.1218",
    pages = "23875--23889",
    ISBN = "979-8-89176-332-6"
}

@article{Li2024PatchFinderAT,
  title={PatchFinder: A Two-Phase Approach to Security Patch Tracing for Disclosed Vulnerabilities in Open-Source Software},
  author={Kai-Jing Li and Jian Zhang and Sen Chen and Han Liu and Yang Liu and Yixiang Chen},
  journal={Proceedings of the 33rd ACM SIGSOFT International Symposium on Software Testing and Analysis},
  year={2024},
  url={https://api.semanticscholar.org/CorpusID:271404420}
}

\newpage

\appendix

\section{Top-Ranked Function Call Chains}
\label{app:chains}
Top 80 (3\%) of 2,713 ranked function call chains. Functions in the top-ranked
cluster are highlighted in orange (contains vulnerability fix), yellow
(authentication-related), and gray (logging-related). \bigskip

\begin{adjustwidth}{-0.5cm}{-0.5cm}
	\centering

	{\centering
		\begin{minipage}[t]{0.48\linewidth}
			\vspace{0pt}

			\csvreader[
				separator=pipe,
				tabular={ccp{5cm}},
				table head=\toprule Rank & Iterations & Call Chains \\\midrule,
				late after line=\\,
				table foot=\bottomrule,
				filter expr={ test{\ifnumless{\thecsvinputline}{42}} }
			]{chains.psv}{rank=\rank,rounds=\rounds,value=\desc}
			{
				\rank &
				\rounds &
				\desc
			}

		\end{minipage}
		\hfill
		\begin{minipage}[t]{0.48\linewidth}
			\vspace{0pt}

			\csvreader[
				separator=pipe,
				tabular={ccp{5cm}},
				table head=\toprule Rank & Iterations & Call Chains \\\midrule,
				late after line=\\,
				table foot=\bottomrule,
				filter expr={ test{\ifnumgreater{\thecsvinputline}{41}} and test{\ifnumless{\thecsvinputline}{82}} }
			]{chains.psv}{rank=\rank,rounds=\rounds,value=\desc}
			{\rank & \rounds & \texttt{\desc}}
		\end{minipage}
	}

\end{adjustwidth}

\newpage

\section{Top-Ranked Function Clusters}
\label{app:clusters}

Top 35 (30\%) of 119 ranked function call clusters. Clusters that contain
functions in top-ranked cluster are highlighted in orange (contains
vulnerability fix), yellow (authentication-related), and gray
(logging-related). \bigskip

\begin{table}[h]
	\centering
	\begin{filecontents*}{clusters.csv}
rank,seed,diam,size,mass,density,score,priority
1,\hlorange{\texttt{sub\_2aac220}},2,5,14.8333,2.9667,44.0056,1
2,\hl{\texttt{sub\_2aaab40}},2,3,10.0500,3.3500,33.6675,2
3,\hl{\texttt{sub\_2ab01f0}},1,3,8.8333,2.9444,26.0093,2
4,\hl{\texttt{sub\_2659750}},2,9,11.0956,1.2328,13.6792,2
5,\hl{\texttt{sub\_26f0d30}},1,3,4.4048,1.4683,6.4673,2
6,\hl{\texttt{sub\_2ac4ed0}},3,25,12.0917,0.4837,5.8483,2
7,\hlorange{\texttt{sub\_2aac220}},3,91,20.0338,0.2202,4.4105,1
8,\hl{\texttt{sub\_2659750}},1,3,3.5417,1.1806,4.1811,2
9,\hl{\texttt{sub\_2cba350}},1,3,3.3442,1.1147,3.7278,2
10,\hl{\texttt{sub\_2ac4ed0}},2,7,4.6637,0.6662,3.1072,2
11,\hl{\texttt{sub\_2abd080}},2,52,12.1591,0.2338,2.8431,2
12,\hl{\texttt{sub\_2394810}},3,95,13.5400,0.1425,1.9298,2
13,\hl{\texttt{sub\_23595d0}},3,88,12.7988,0.1454,1.8615,2
14,\hl{\texttt{sub\_230e110}},2,84,12.4930,0.1487,1.8580,2
15,\hl{\texttt{sub\_2002510}},3,91,12.9823,0.1427,1.8521,2
16,\hl{\texttt{sub\_2cffa30}},3,91,12.9816,0.1427,1.8519,2
17,\hl{\texttt{sub\_2a34360}},3,85,12.5091,0.1472,1.8409,2
18,\hl{\texttt{sub\_2921070}},3,85,12.5054,0.1471,1.8398,2
19,\hl{\texttt{sub\_32e7ea0}},3,85,12.5051,0.1471,1.8397,2
20,\hl{\texttt{sub\_32fcfc0}},3,85,12.5010,0.1471,1.8385,2
21,\hl{\texttt{sub\_1db5e80}},3,101,13.6186,0.1348,1.8363,2
22,\hl{\texttt{sub\_1dffba0}},3,86,12.5188,0.1456,1.8223,2
23,\hl{\texttt{sub\_2ad2ff0}},3,88,12.6559,0.1438,1.8201,2
24,\hl{\texttt{sub\_2cecf80}},3,101,13.5564,0.1342,1.8196,2
25,\hl{\texttt{sub\_2da0950}},3,94,13.0633,0.1390,1.8154,2
26,\hl{\texttt{sub\_2aa3940}},2,19,5.8525,0.3080,1.8027,2
27,\hlgray{\texttt{sub\_2aae970}},1,3,2.1048,0.7016,1.4767,
28,\hlgray{\texttt{sub\_2d5c920}},1,3,2.0201,0.6734,1.3603,
29,\hlgray{\texttt{sub\_2d7e960}},1,3,1.9473,0.6491,1.2640,
30,\hlgray{\texttt{sub\_236b990}},1,3,1.9365,0.6455,1.2500,
31,\hl{\texttt{sub\_2b51b30}},2,36,6.6972,0.1860,1.2459,2
32,\hlgray{\texttt{sub\_22a9630}},1,3,1.9196,0.6399,1.2283,
33,\hlgray{\texttt{sub\_212ded0}},1,3,1.9173,0.6391,1.2254,
34,\hlgray{\texttt{sub\_1d408b0}},1,3,1.9128,0.6376,1.2196,
35,\hl{\texttt{sub\_2d71f50}},3,61,8.0052,0.1312,1.0505,2
\end{filecontents*}
\csvreader[
		separator=comma,
		tabular={cccccccc},
		table head=\toprule Rank & Seed & Diameter & Size & Mass & Density & Score \\\midrule,
		late after line=\\,
		table foot=\bottomrule
	]{clusters.csv}{rank=\rank,seed=\seed,diam=\diam,size=\size,mass=\mass,density=\density,score=\score}
	{\rank & \seed & \diam & \size & \mass & \density & \score}
\end{table}

\end{document}